\documentclass[aps,prd,twocolumn,groupedaddress,showpacs,amsmath,amssymb]{revtex4}
    \newcommand{\be}[1]{\begin{equation}\label{#1}}
    \newcommand{\ep}[1]{\epsilon_{#1}}
    \newcommand{\de}[1]{\delta_{#1}}

    \newcommand{\rd}{{\rm d}}
    \newcommand{\ds}{{\delta\sigma}}
    \newcommand{\re}{{\rm e}}
    \newcommand{\pa}[1]{\left(#1\right)}
    \newcommand{\paq}[1]{\left[#1\right]}
    
    \newcommand{\av}[1]{\langle#1\rangle}
    \newcommand{\M}{{\rm M_{\rm P}}}
    \def\ee{\end{equation}}
    \def\ba{\begin{eqnarray}}
    \def\ea{\end{eqnarray}}

\usepackage{graphicx,latexsym}
\usepackage{graphicx,epsf}
\usepackage{color}
\usepackage{epsfig}
\begin{document}
\title{Inflation and Reheating in Spontaneously Generated Gravity}
\author{A. Cerioni$\,^{1,2}$, F. Finelli$\,^{3,2}$, A. 
Tronconi$\,^{1,2}$ and G. Venturi$\,^{1,2}$}

\affiliation{$^{1}$ Dipartimento di Fisica, Universit\`a degli Studi di
Bologna, via Irnerio, 46 -- I-40126 Bologna -- Italy}
\affiliation{$^{2}$ INFN, Sezione di Bologna,
Via Irnerio 46, I-40126 Bologna, Italy}
\affiliation{
$^{3}$ INAF/IASF Bologna,
Istituto di Astrofisica Spaziale e Fisica
Cosmica di Bologna \\
via Gobetti 101, I-40129 Bologna - Italy}
%%%%%%%%%%%%
\begin{abstract}
Inflation is studied in the context of induced gravity (IG) $\gamma \sigma^2 R$, where $R$ is the Ricci scalar, $\sigma$ a scalar field and $\gamma$ a dimensionless constant, and diverse symmetry-breaking potentials $V(\sigma)$ are considered. In particular we compared the predictions for Landau-Ginzburg (LG) and Coleman-Weinberg (CW) type potentials and their possible generalizations with the most recent data. We find that large field inflation generally leads to fewer constraints on the parameters and the shape of the potential whereas small field inflation is more problematic and, if viable, implies more constraints, in particular on the parameter $\gamma$. We also examined the reheating phase and obtained an accurate analytical solution for the dynamics of inflaton and the Hubble parameter by using a multiple scale analysis (MSA). The solutions were then used to study the average expansion of the Universe, the average equation of state for the scalar field and both the perturbative and resonant decays of the inflaton field.
\end{abstract}
\maketitle
\section{Introduction}
Many years ago a model for a varying gravitational coupling was introduced \cite{brans}. The model consisted of a massless scalar field whose inverse was associated with the gravitational coupling. Such a field evolved dynamically in the presence of matter and led to cosmological predictions differing from Einstein Gravity (EG) in that one generally obtained a power-law time dependence for the gravitational coupling. Subsequently it was suggested that the gravitational constant is generated as a one loop effect in some fundamental interaction \cite{sakharov} or through spontaneous symmetry breaking (for reviews see \cite{Zee:1980sj,adler}). In order to reduce the strong time dependence in a cosmological setting, which remained after the introduction of matter, a simple globally scale invariant model for induced gravity (IG) involving a scalar field $\sigma$ and a quartic potential $\lambda \sigma^{4}$ was introduced \cite{CV}. The spontaneous breaking of scale invariance in such a context, either through the presence of a condensate \cite{zee,Accetta,Sorbo} or quantum effects \cite{CW,spokoiny}, then led to EG plus a cosmological constant and, on treating matter as a perturbation, a time dependence for the scalar field (gravitational constant) and consistent results \cite{CV}. A more detailed analysis \cite{FTV} of such a simple model including both radiation and matter showed that it led to EG plus a cosmological constant as a stable attractor among homogeneous cosmologies and was therefore a viable Dark Energy model for a range of scalar field initial conditions and a positive $\gamma$ coupling to the Ricci scalar $\gamma\sigma^{2}R$. In that earlier study we considered values for the scalar field which were sufficiently close to the spontaneously broken symmetry equilibrium values for the scalar field and compared our results with present values of the cosmological constant and the solar system data.
In a later note \cite{CFTV} we studied the above approach for values of our parameters sufficiently far from the equilibrium values (back in time) for sufficient inflation and the subsequent reheating to take place and examined the compatibility of the predictions for different symmetry breaking potentials with the current data.\\
The aim of this paper is to investigate in detail the dynamics of the scalar field and the generation of scalar and tensor perturbations in the early Universe.\\
The paper is organized as follows: in section II the formalism and the general equations for the homogeneous dynamics are described. In section III we formalize the problem of perturbations in the IG context and section IV is dedicated to comparing the predictions of diverse symmetry-breaking potentials with the most recent data. In section V we apply a Multiple Scale Analysis (MSA) to the phase of coherent oscillations of the inflaton field at the end of inflation. In section VI we study the perturbative decay of the inflaton field in the IG context and in section VII the resonant decay of the inflaton is discussed. Finally, in section VIII, our conclusions are summarized.
%%%%%%%%%%%%%%%%%%%%%%%%%%%%%%%%%
\section{IG inflation}
We consider the system described by the action
\begin{equation}
S = \int d^4 x \sqrt{-g}\left[-\frac{g^{\mu\nu}}{2}\partial_{\mu}
\sigma\partial_{\nu}\sigma+{\gamma\over 2}\sigma^2 R-V(\sigma) \right]
\label{original}
\end{equation}
where $\gamma$ is a dimensionless, positive definite parameter giving the non-minimal coupling between the scalar field and gravity. The Einstein-Hillbert term for gravity is replaced by an effective mass-like term for the scalar field $\sigma$. In turn such a scalar field is also responsible for the existence of space-time curvature and its expectation value dynamically generates an effective Planck mass. If we restrict our analysis to the homogeneous dynamics and we assume a spatially flat Robertson-Walker background 
\be{metric}
d s^2 = g_{\mu \nu} dx^{\mu} dx^{\nu}=- dt^2+a^2(t) d\vec{x}^2 \,,
\ee
the variation of the above Lagrangian leads to the following set of 
independent equations
\begin{eqnarray}
H^2 &=& \frac{1}{3\gamma \sigma^2}
\left[ \frac{\dot\sigma^2}{2}+V(\sigma) \right]-
2H\frac{\dot\sigma}{\sigma}\label{FrEQ}\\
\ddot\sigma &+& 3H\dot\sigma + \frac{\dot\sigma^2}{\sigma}= - 
\frac{V_{\mathrm{eff},\sigma}}{1+6\gamma}
\label{KGEQ}
\end{eqnarray}
where we defined $V_{\mathrm{eff},\sigma}=\rd V/\rd \sigma-4V/\sigma$. 
The l.h.s. of Eq.\ (\ref{FrEQ}) is positive definite or zero whereas the l.h.s of the same equation can also be negative in some regions of the phase-space. As a consequence one has some restrictions on the allowed phase-space.
In particular one needs
\be{PSbound}
6\gamma-\sqrt{6\gamma\left(1+6\gamma\right)}<\frac{\dot \sigma}{H\sigma}<6\gamma+\sqrt{6\gamma\left(1+6\gamma\right)}
\ee
in order for (\ref{FrEQ}) to be well defined.

It is known \cite{Maeda,Flanagan} that by a conformal transformation:
\ba
{\tilde g}_{\mu \nu} &=& \Omega^2 g_{\mu \nu} \nonumber \\
%{\tilde \phi} - {\tilde \phi_0} &=& M_{\rm pl} \sqrt{\frac{1+6\gamma}{\gamma}}
%\log \left( \frac{\phi}{\phi_0}\right) \nonumber \\
d {\tilde \sigma}^2 & =&  \frac{\left( 1 + 6 \gamma \right)}{\Omega^2} d \sigma^2 \\
{\tilde V}&=& \Omega^{-4} V \,.\nonumber
%{\tilde g}_{\mu \nu} &=& \frac{\gamma \phi^2}{M_{\rm pl}^2}  g_{\mu \nu} \nonumber \\
%{\tilde \phi} - {\tilde \phi_0} &=& M_{\rm pl} \sqrt{\frac{1+6\gamma}{\gamma}}
%\log \left( \frac{\phi}{\phi_0}\right) \nonumber \\
%{\tilde V} ({\tilde \phi}) &=& \gamma^2 \phi^4 V(\phi) \,.
\ea
where $\Omega^2 = \gamma \sigma^2/\M^2$, we can rewrite (up to a boundary term) the action in Eq. (\ref{original}) as:
%leaves the action invariant up to a boundary term.
\be{EinsteinFrame}
S_E = \int d^4 x \sqrt{-{\tilde g}} \left[- 
\frac{{\tilde g}^{\mu \nu}}{2} \partial_\mu {\tilde \sigma} \partial_\nu {\tilde \sigma} +\frac{{\tilde R} \, \M^2}{2}-
{\tilde V}({\tilde \sigma}) \right]
\ee
and we have introduced a (reduced) Planck mass $\M$ ($=(8 \pi G)^{-1/2}$ with $G$ the 
Newton constant).

The variation of Eq.\ (\ref{EinsteinFrame}) then leads to a set of equations analogous to those obtained from Eq.\ (\ref{original}).
Let us note that the spectrum of 
curvature perturbations and the amplitude 
of gravitational waves obtained are invariant under conformal transformations \cite{Chiba}: it is for this reason that
inflationary calculations are often performed in the Einstein frame.\\
We observe that other 
important quantities in cosmology are not left invariant under 
conformal transformation. This is the case for the Hubble parameter $H$. If we are interested in late 
time cosmology and use observational data to constrain $H$ in a scalar-tensor theory \cite{FTV}, this should be 
done for the Hubble parameter in the Jordan frame, which is different from the Hubble parameter in the Einstein frame. To conclude, for us, the Jordan frame is the physical one.

%%%%%%%%%%%%%%%%%%%%%%

\subsection{Hubble and scalar field flow functions}

As emphasized in \cite{CFTV}, cosmological linear perturbations depend not only on the derivatives of the Hubble parameter, 
but also on the derivative of the scalar field itself. 
It is therefore useful to introduce the hierarchy of scalar field flow functions $\delta_n$ ($d\ln |\delta_n|/dN \equiv \delta_{n+1}$ with
$n \ge 0$, $\delta_0 \equiv \sigma/\sigma(t_{i})$) in addition to $\epsilon_n$ ($d\ln |\epsilon_n|/dN \equiv \epsilon_{n+1}$, 
$\epsilon_0 \equiv H(t_{i})/H$) with $n \ge 0$, where
$t_{i}$ is some initial time and $N\equiv\ln \frac{a}{a(t_{i})}$
is the number of e-folds. These two hierarchies are related by:
\be{eom1}
\epsilon_1 = \frac{\delta_1}{1+\delta_1}
\left(\frac{\delta_1}{2\gamma}+2\delta_1+\delta_2-1\right).
\ee
 
The above parameters are related to the Hubble flow functions in the Einstein frame ($\tilde \epsilon_{i}$) by 

\ba
{\tilde \epsilon}_1 &=&  \frac{\left(1+6\gamma\right)\de{1}^{2}}{2\gamma\left(1+\de{1}\right)^{2}}\nonumber \\ 
{\tilde \epsilon}_2 &=& \frac{2 \delta_2}{(1+\delta_1)^2}.
\ea
These hierarchies arise naturally with the slow-roll (SR) of the inflaton field.  Note that in EG the equations governing the dynamics of the scalar and tensor fluctuations during inflation can be written in terms of the Hubble flow function hierarchy. In IG the equivalent set of equations cannot be written only in terms of Hubble flow function hierarchy and the scalar field flow function hierarchy is also needed.

\subsection{Homogeneous Dynamics as flow of the Hubble and Scalar Field Flow Functions}
\begin{figure}[t!]
\centering
\epsfig{file=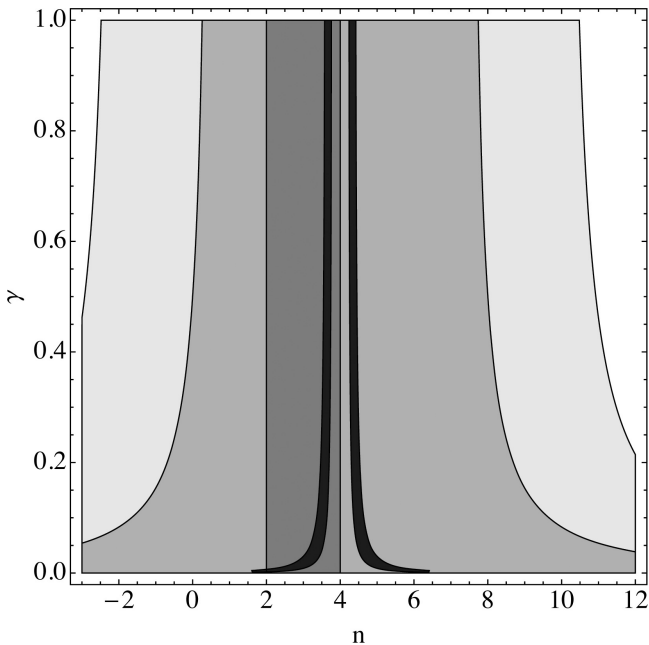, width=8.5 cm}
\caption{{\it The figure represents $\ep{1}$ in (\ref{exsol}) as a function of $n$ and $\gamma$.
The lighter grey region is for $\ep{1}>1$, the darker grey region is for $\ep{1}<0$ and the
intermediate grey region is for $0<\ep{1}<1$. The lines for $n=2,4$ represents the de Sitter solutions. The darkest grey areas are the domains allowed by the present experimental constraints (\ref{obscons}).}
\label{ESB}}
\end{figure}

The homogeneous dynamics of the field-gravity system and the slow-roll conditions for inflation have 
a peculiar role in the theory of cosmological perturbations since they provide an approximate 
method to determine the dynamics of these perturbations and compare theoretical models with observations.
Exact solutions for such a dynamics can be also found for particular choices of the inflaton potential 
both in the EG and IG framework.
On writing the equations of motion in terms of the SR parameters $\ep{i}$ and $\de{i}$ one can easily find a set of 
these solutions. In EG one has
\be{dehierEG}
\tilde\delta_{1}^{2}\,\frac{\tilde\sigma^{2}}{\M^{2}}=\tilde\epsilon_{1}\,\Rightarrow\,\tilde\delta_{1}+\tilde\delta_{2}=\frac{\tilde\epsilon_{2}}{2}\,,
\ee
and the Klein-Gordon equation for the scalar field can be rewritten as
\be{KGEG}
\tilde\delta_{2}+\tilde\delta_{1}-\tilde\epsilon_{1}+3+\frac{\tilde\delta_{1}}{\tilde\epsilon_{1}}\frac{\rd \ln \tilde V}{\rd\ln\tilde\sigma}\pa{3-\tilde\epsilon_{1}}=0.
\ee
From Eqs.\ (\ref{dehierEG},\ref{KGEG}) one easily observes that no solution with $\tilde\delta_{1}$ 
and $\tilde\epsilon_{1}$ simultaneously constant and different from zero exists while a non trivial solution 
can be found for the case $\tilde\epsilon_{2}=0$, $\tilde\delta_{1}=\pm\sqrt{\tilde\epsilon_{1}}\,\M/\tilde\sigma$ and 
$\tilde V\propto \exp\pa{\sqrt{\tilde\epsilon_{1}} \tilde\sigma/\M}$, namely the well-know power-law inflationary solution.\\
In the IG context Eqs.\ (\ref{dehierEG},\ref{KGEG}) are replaced by Eq. (\ref{eom1}) and
\begin{eqnarray}
\ep{1}& = &\left[3\left(\de{1}-4\gamma+\gamma \frac{\rd \ln V}{\rd \ln \sigma}\right)+\de{1}
\left(6\gamma-\frac{\de{1}}{2}\right)\frac{\rd \ln V}{\rd \ln\sigma}\right.\nonumber\\
&&\left.\phantom{\frac{A}{B}}\!\!\!\!\!\!\!+\de{1}\left(\de{1}+\de{2}\right)\right]\times
\frac{1}{\de{1}-6\gamma}\,.\label{eom2}
\end{eqnarray}
Despite their quite involved form one can still see that an exact, non trivial solution exists 
with $\ep{2}=\de{2}=0$ when $V=V_{0}\pa{\sigma/\sigma_{0}}^{n}$, that is
\be{exsol}
\de{1}=-\frac{\gamma\pa{n-4}}{1+\gamma\pa{n+2}},\;\;\ep{1}=\frac{\gamma\pa{n-2}\pa{n-4}}{2+2\gamma\pa{n+2}}.
\ee
On solving Eqs.\ (\ref{eom1},\ref{eom2}) one obtains
\be{nonacceptsol}
\de{1}=6\gamma\pm\sqrt{6\gamma\left(1+6\gamma\right)},\quad \ep{1}=3+2\de{1}
\ee
which are not compatible with phase-space constraints. In Fig. (\ref{ESB}) we plotted the
behavior of $\ep{1}$ in (\ref{exsol}) on varying $\gamma$ and $n$.
Decelerated, accelerated and super-accelerated solutions can be found inside the boundaries of
phase space as can be seen in Fig.\ (\ref{ESB}) where the lighter region represents decelerated
trajectories ($\ep{1}>1$), the darker area is for super-accelerated solutions ($\ep{1}<0$) and
the intermediate grey region represents the accelerated trajectories ($0<\ep{1}<1$). The outer
boundaries of the lighter gray area are those of phase space and are also those of
the region of stability for such exact solutions with respect to homogeneous perturbations. The lines $n=2$ and $n=4$
dividing accelerated and super-accelerated trajectories represents the De Sitter solutions. Note finally
that the exact solutions with $n\ge4$ have $\de{1}\le 0$ (Large Field dynamics) while for $n<4$ one has $\de{1}>0$
(Small Field dynamics).

%%%%%%%%%%%%%%%%
\section{Cosmological Perturbations}

Scalar curvature perturbations produced by quantum fluctuations of the
inflaton during the accelerated stage are
described by $\mathcal R(x)= - H \delta \sigma(x)/\dot \sigma$ in the uniform curvature gauge \cite{hwang},
where $\delta \sigma(x)$ is the scalar inflaton perturbation and is the correct field variable
to quantize.
The Fourier component $\delta \sigma_{k}$ ($\mathcal{R}_{k}=-H\delta \sigma_{k}/\dot \sigma$) of the inflaton fluctuation,
in the IG context, has been shown to satisfy the following differential equation
\cite{hwang}:
\ba
\ddot{\delta \sigma_k}&+&\left( 3 H + \frac{\dot Z}{Z} \right)
\dot{\delta \sigma_k} + \nonumber\\
&+&\left[ \frac{k^2}{a^2} - \frac{1}{a^3 Z \sigma \delta_1}
\left( a^3 Z \left( \sigma \delta_1 \right)^\cdot \right)^\cdot
\right] \delta \sigma_k = 0\label{scperteq}
\ea
where
\be{Qshom}
Z = \frac{H^2 \sigma^2 (1+6\gamma)}{(\dot \sigma + H \sigma)^2} = \frac{1+6\gamma}{(1 + \delta_1)^2}.
\ee
The above equation can be rewritten for $S_k \equiv a \sqrt{Z}\,\delta\sigma_k$ as:
\be{scalarnew}
\frac{\rd^2 S_k}{\rd\eta^2} +\left[k^2 +M_{S}^{2}(\eta)\right]S_k =0
\ee
where
\begin{widetext}
\begin{equation} 
M_{S}^{2}(\eta)\equiv-\mathcal{H}^2\left[\delta_1^2+\delta_2^2+(3-\epsilon_1)(\delta_1+\delta_2+1)+\delta_2\delta_3
+\frac{\delta_1\delta_2}{1+\delta_1}\left(\epsilon_1+\delta_1-
3\delta_2-\delta_3+\frac{2\delta_1\delta_2}{1+\delta_1}-2\right)-1\right]\,,
\end{equation}
\end{widetext}
with $\eta$ the conformal time ($a(\eta)\rd \eta\equiv\rd t$) and $\mathcal H\equiv a^{-1}\rd a/\rd \eta$.
Gravitational waves are also produced during inflation. 
In IG the Fourier modes of tensor perturbations satisfy the following equation:
\be{tnperteqN}
\ddot h_{s,k} + (3H + 2H\delta_1)\dot h_{s,k}+\frac{k^2}{a^2}h_{s,k}=0
\ee
where $s = + \,, \times$ denotes the two polarization states.
On setting $T_{s,k} \equiv \frac{1}{\sqrt{2}}a \sigma \sqrt{\gamma}\,h_{s,k}$ the above equation can be rewritten as:
\begin{equation}
\frac{\rd^{2}T_{s,k}}{\rd \eta^{2}}+\left[k^2+M_{T}^{2}(\eta)\right]T_{s,k}=0\,,
\end{equation}
where
\be{tensmassnew}
M_{T}^{2}(\eta)\equiv -\mathcal{H}^2\left[2-\epsilon_1+\delta_1(3+\delta_1+\delta_2-\epsilon_1)\right].
\ee
We define the power spectra of scalar curvature perturbations
and tensor perturbation as
\be{scpowspe}
\mathcal{P}_{\mathcal{R}}(k) \equiv \frac{k^3}{2\pi^2}|\mathcal{R}_k|^2\simeq \mathcal{P}_{\mathcal{R}}(k_*)
\left( \frac{k}{k_*} \right)^{n_s-1}
\ee
and
\be{tnpowspe}
\mathcal{P}_{h}(k) \equiv \frac{2 k^3}{\pi^2}\left(|h_{+ , k}|^2
+ |h_{\times , k}|^2 \right) \,\simeq \mathcal{P}_{h}(k_*)
\left( \frac{k}{k_*} \right)^{n_t}
\ee
respectively, where $k_*$ is a suitable pivot scale.

\subsection{Exact Solutions for $V(\sigma) \propto \sigma^n$}

In IG power-law potentials for $\sigma$ lead to power-law inflation \cite{CFTV} and the spectral index for such background solutions is:
\be{exspecind}
n_{s}-1=n_{t}=\frac{2\gamma \pa{n-4}^{2}}{\gamma\pa{n-4}^{2}-2\pa{6\gamma+1}}
\ee 
and the exact tensor-to-scalar ratio is given by 
\be{consrel}
r=\frac{\mathcal{P}_{h}(k)}{\mathcal{P}_{\mathcal{R}}(k)}=
-\frac{8\,n_{t}}{1-\frac{n_{t}}{2}}
\ee
which agrees with the consistency condition of power-law inflation in EG. 

From the most recent compilation of data \cite{fhll} we obtain the constraint:
\be{obscons}
39 < \frac{6 \gamma +1}{\gamma (n-4)^2} < 123
\ee
at the 95 $\%$ confidence level which is independent of $\gamma$ for $\gamma$ large.

%%%%%%%%%%%%%%%%%%%%%%%%%%%%%%

\subsection{Slow Roll Results}

For generic potentials the form of the exact solutions 
of Eqs.\ (\ref{eom1},\ref{eom2}) is unknown 
and one must employ the SR approximation 
in order to obtain analytical estimates for the spectra of perturbations. In such an approximation
Eqs. (\ref{FrEQ},\ref{KGEQ}) become
\begin{eqnarray}
H^2 &\simeq& \frac{V(\sigma)}{3\gamma \sigma^2}
\label{FrEQSR}\\
3H\dot\sigma &\simeq& - 
\frac{V_{\mathrm{eff},\sigma}}{1+6\gamma}.
\label{KGEQSR}
\end{eqnarray}
On assuming $\de{i}\ll 1$ (and consequently $\ep{i}\ll1$) 
one can thus neglect their variation in time and rely, 
to first order in the SR parameters, on the results 
obtained for exact evolution with $\de{2}\sim\de{1}\neq 0$. 
Such an approach leads to the following expressions for the spectra:
\be{scspecSR}
\mathcal{P}_{\mathcal{R}}(k_*) \simeq
\frac{A \, H^{2}_*}{4 \pi^{2}\pa{1+6\gamma}\de{1 \, *}^{2}\sigma_*^{2}}
\ee
where
\be{Acoeff}
A = \paq{1-2\ep{1\,*}+C \pa{\de{1\,*}+\de{2\,*}+\ep{1\,*}}},
\ee
$C=2\pa{2-\ln 2-b}$, $b$ is the Euler-Mascheroni constant and
\be{tnspecSR}
\mathcal{P}_{h}(k_*) \simeq
\frac{2 (A - C \delta_{2 \, *}) H_*^{2}}{\pi^{2}\gamma\sigma_*^{2}}\,.
\ee
The spectral indices for the perturbations are
\be{indexesSR}
n_{s}-1=n_{t}-2\de{2}=-2\pa{\de{1}+\de{2}+\ep{1}}\,.
\ee
and the tensor to scalar ratio is $r=-8\,n_{t}$.\\
In the SR regime the dynamics of the scalar field can be approximated as follows
\ba
\delta_1 &\equiv&\sigma^{-1}\frac{\rd \sigma}{\rd N}\simeq - \gamma \sigma 
\frac{V_{\mathrm{eff},\sigma}}{(1+6\gamma) \, V} \,,   
\label{delta1}
\\
\delta_2 &\simeq&  - \gamma \sigma^2 
\frac{V_{\mathrm{eff},\sigma\sigma}}{(1+6\gamma) \, V} +
\delta_1 \left( \frac{1+6\gamma}{\gamma}\delta_1 - 3 \right)
\label{delta2}
\ea
and
\be{epapp}
\ep{1}\simeq-\de{1}+\frac{1+6\gamma}{2\gamma}\de{1}^{2}.
\ee
Let us note, from Eqs.\ (\ref{delta1},\ref{delta2}), that the $\de{i}$ are not independent of $\gamma$ and in the $\gamma\ll 1$ limit $\de{1}\sim\gamma$ and thus terms proportional to $\de{1}^{2}/\gamma\sim \de{i}$ should be kept in first order calculations. Second order terms in the above expressions are retained in order to better interpolate the regime from large to small $\gamma$. On using the above expressions and keeping the first order contributions one is finally led to \cite{CFTV}
\be{nsm1pot}
n_{s}-1=\frac{2\gamma\sigma_{*}^{2}}{1+6\gamma}\paq{\frac{V_{\mathrm{eff},\sigma\sigma *}}{V_{*}}-\frac{3V_{\mathrm{eff},\sigma*}}{\sigma_{*}V_{*}}-\frac{3V_{\mathrm{eff},\sigma *}^{2}}{2V_{*}^{2}}}
\ee
and
\be{ntpot}
n_{t}=-\frac{\gamma\sigma_{*}^{2}}{1+6\gamma}\frac{V_{\mathrm{eff},\sigma *}^{2}}{V_{*}^{2}}
\ee
where terms proportional to $V_{\mathrm{eff},\sigma *}^{2}/V_{*}^{2}$ are of first order for $\gamma\ll 1$ and second order for $\gamma\gtrsim 1$.
%%%%%%%%%%%%%%%%%%%%%%%%%%%%%%%%%

\section{Constraints on different potentials}
%\subsection{SR and Observations}

\begin{figure}[t!]
\centering
\epsfig{file=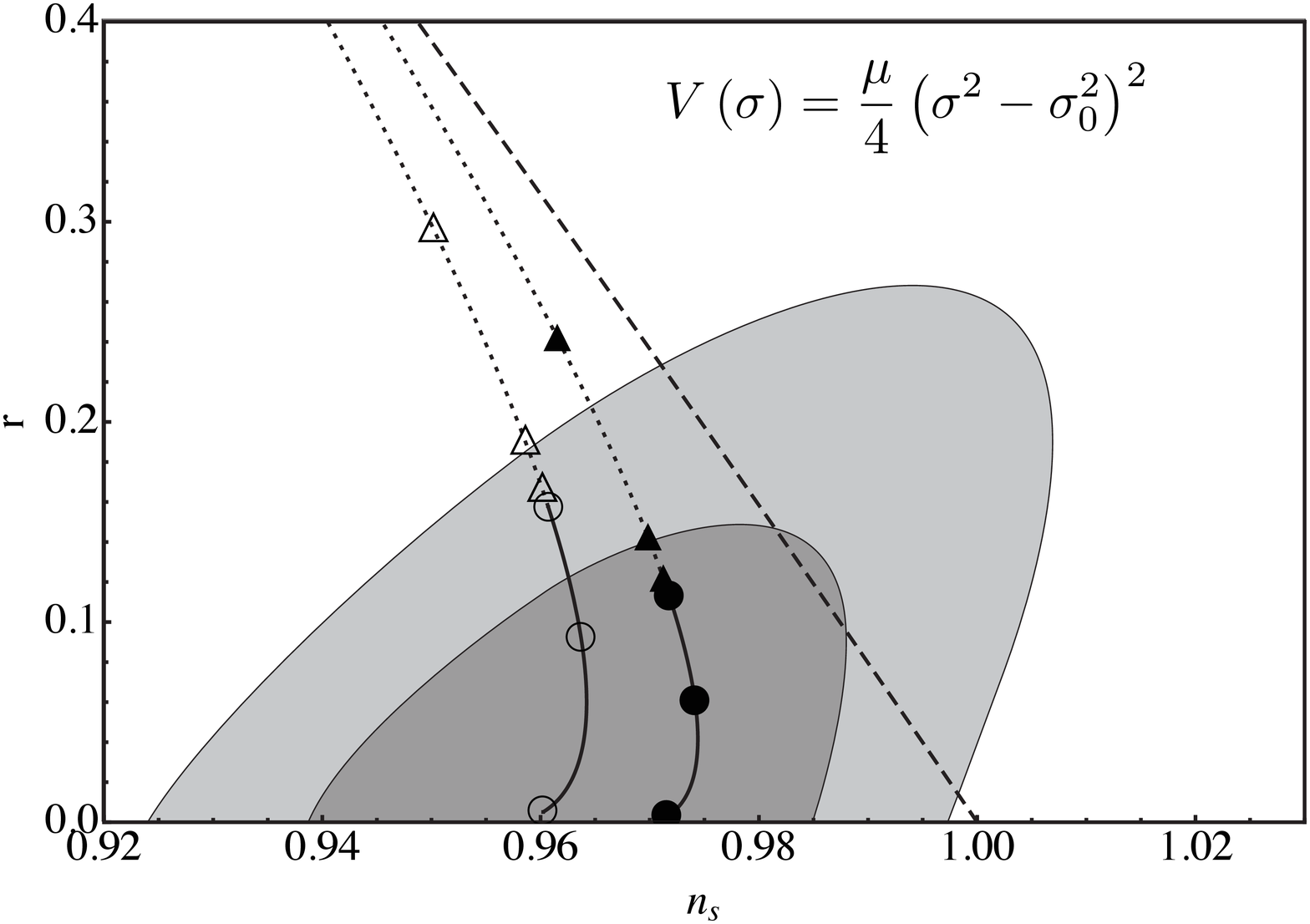, width=8.5 cm}
\epsfig{file=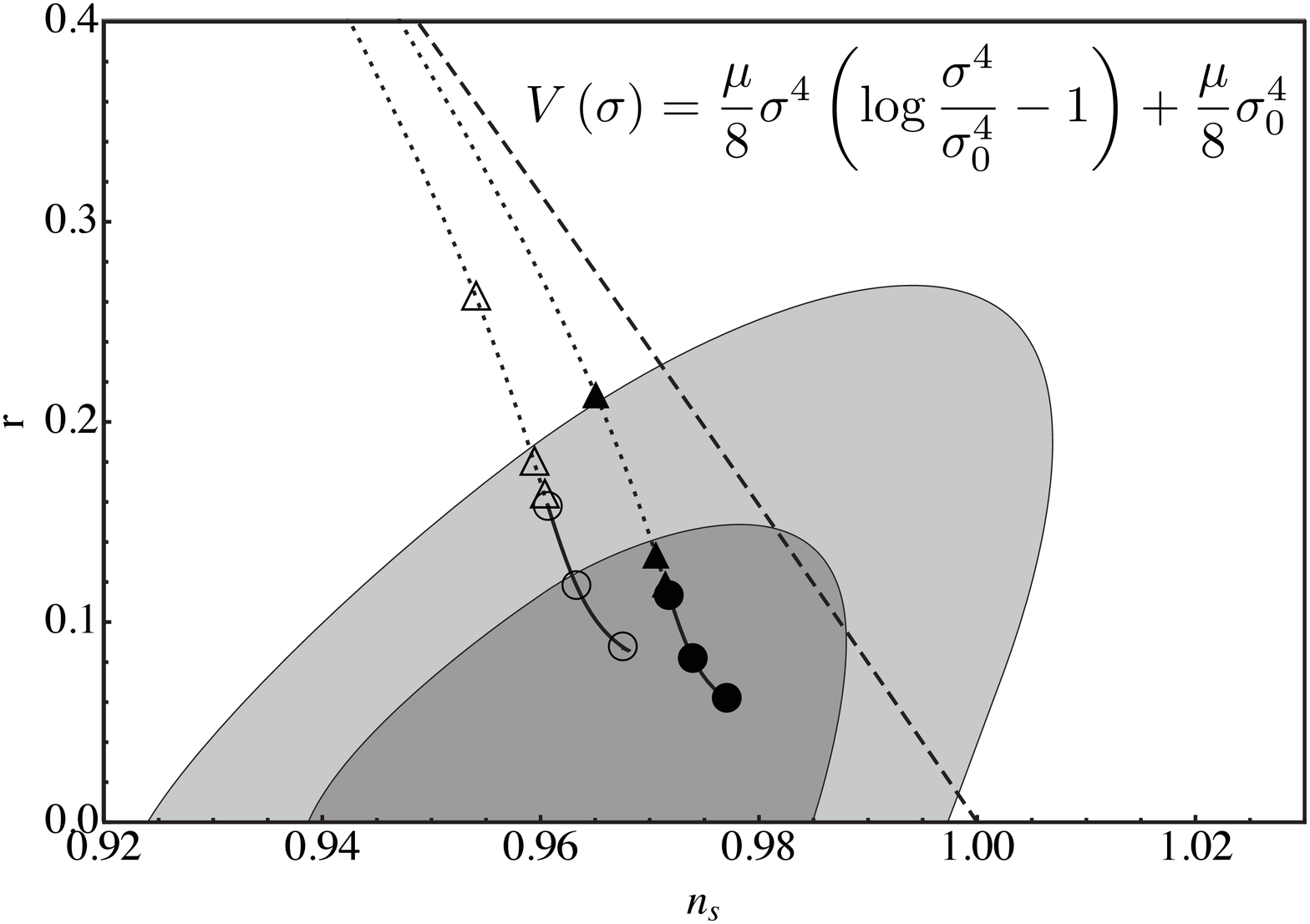, width=8.5 cm}
\epsfig{file=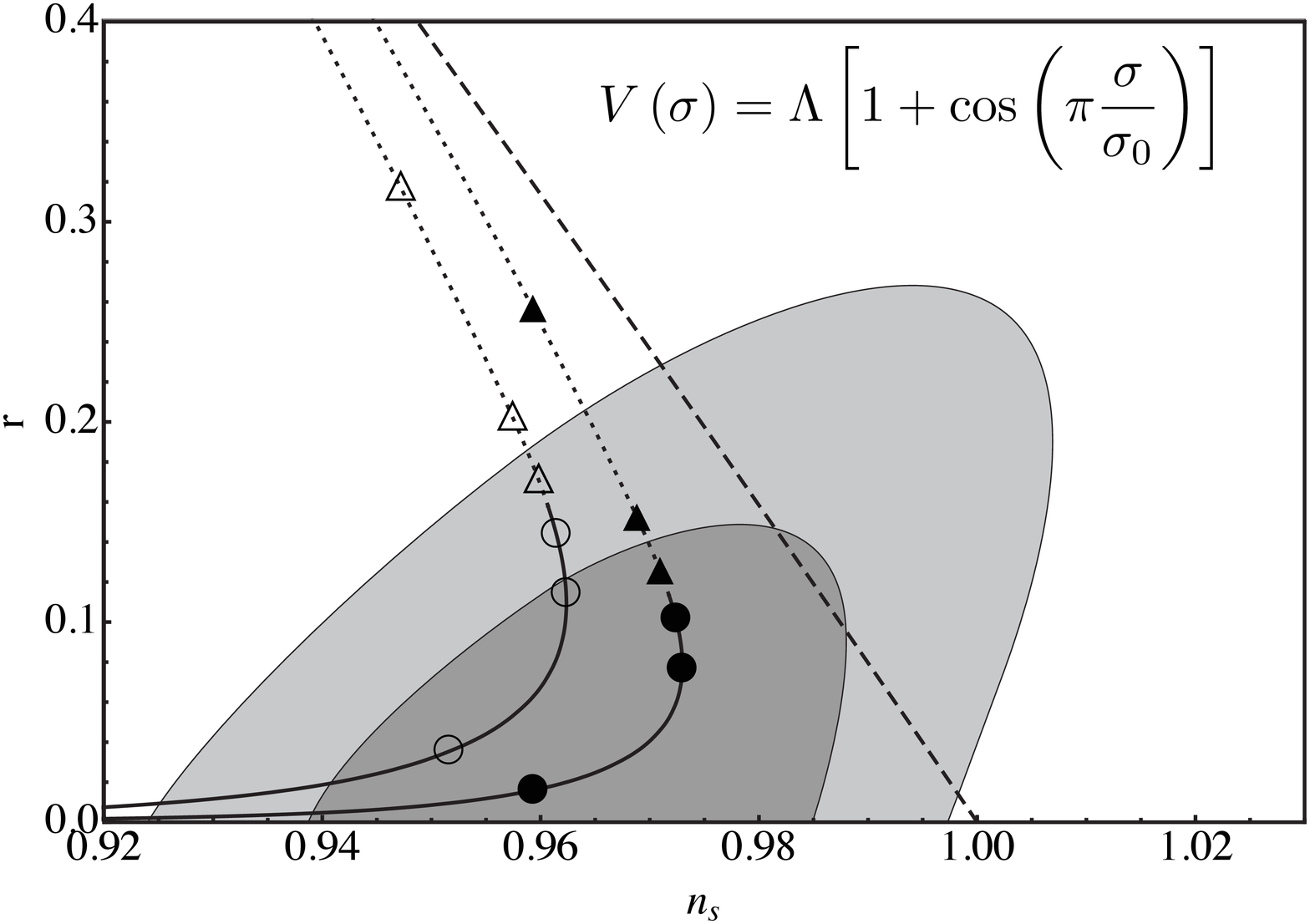, width=8.5 cm}
\caption{{\it The trajectories of the vector $(n_{s},r)$ on varying $\gamma$ and potentials (\ref{LGpot},\ref{CWpot},\ref{NIpot}). For (\ref{LGpot}) and (\ref{CWpot}) the markers represent $L_{\gamma}=-5,-4,-3$ in the SF and $L_{\gamma}=-7,-3,+1$ in the LF regimes. For (\ref{NIpot}) the markers are for $L_{\gamma}=-5,-4,-3$ both in the SF and in the LF regimes. The order of the markers is such that dotted and continuous lines join for $L_{\gamma}\rightarrow-\infty$. The same order is followed in Fig. (\ref{rns2})}
\label{rns1}}
\end{figure}

\begin{figure}[t!]
\centering
\epsfig{file=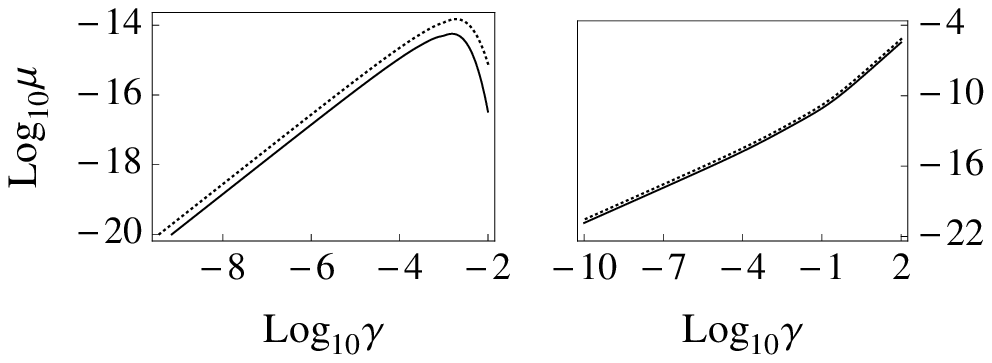, width=8.5 cm}
\epsfig{file=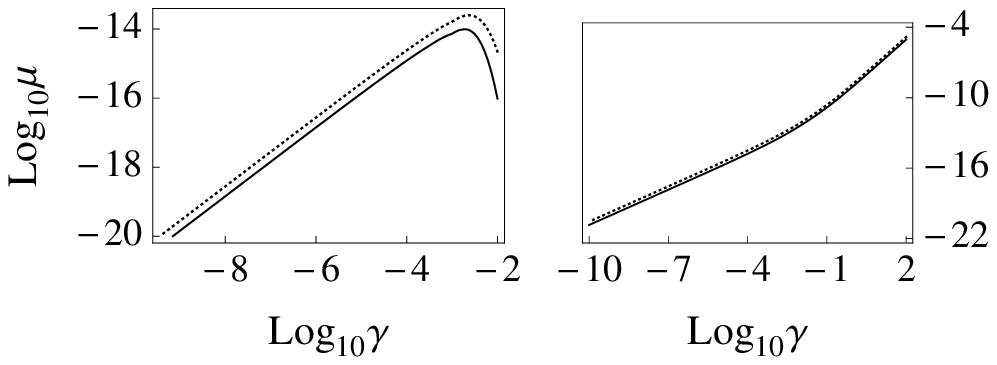, width=8.5 cm}
\epsfig{file=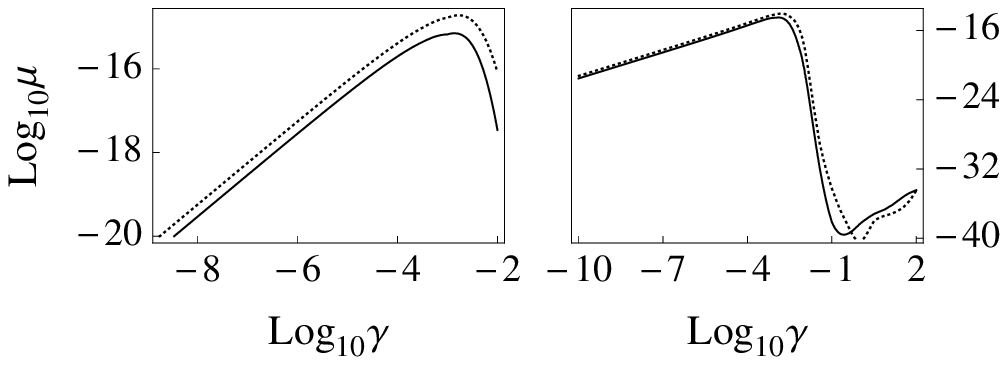, width=8.5 cm}
\caption{{\it Constraints on $\mu$ from the amplitude of the scalar perturbations (\ref{conPR}) for potentials (\ref{LGpot},\ref{CWpot},\ref{NIpot}) respectively, where $\mu\equiv\Lambda/\sigma_{0}^{4}$ for (\ref{NIpot}). The dotted line is for $N_{*}=50$ and the solid line is for $N_{*}=70$ e-folds. The plots on the left refer to the SF case and the ones on the right to the LF.}
\label{PRNI}}
\end{figure}

\begin{figure}[b!]
\centering
\epsfig{file=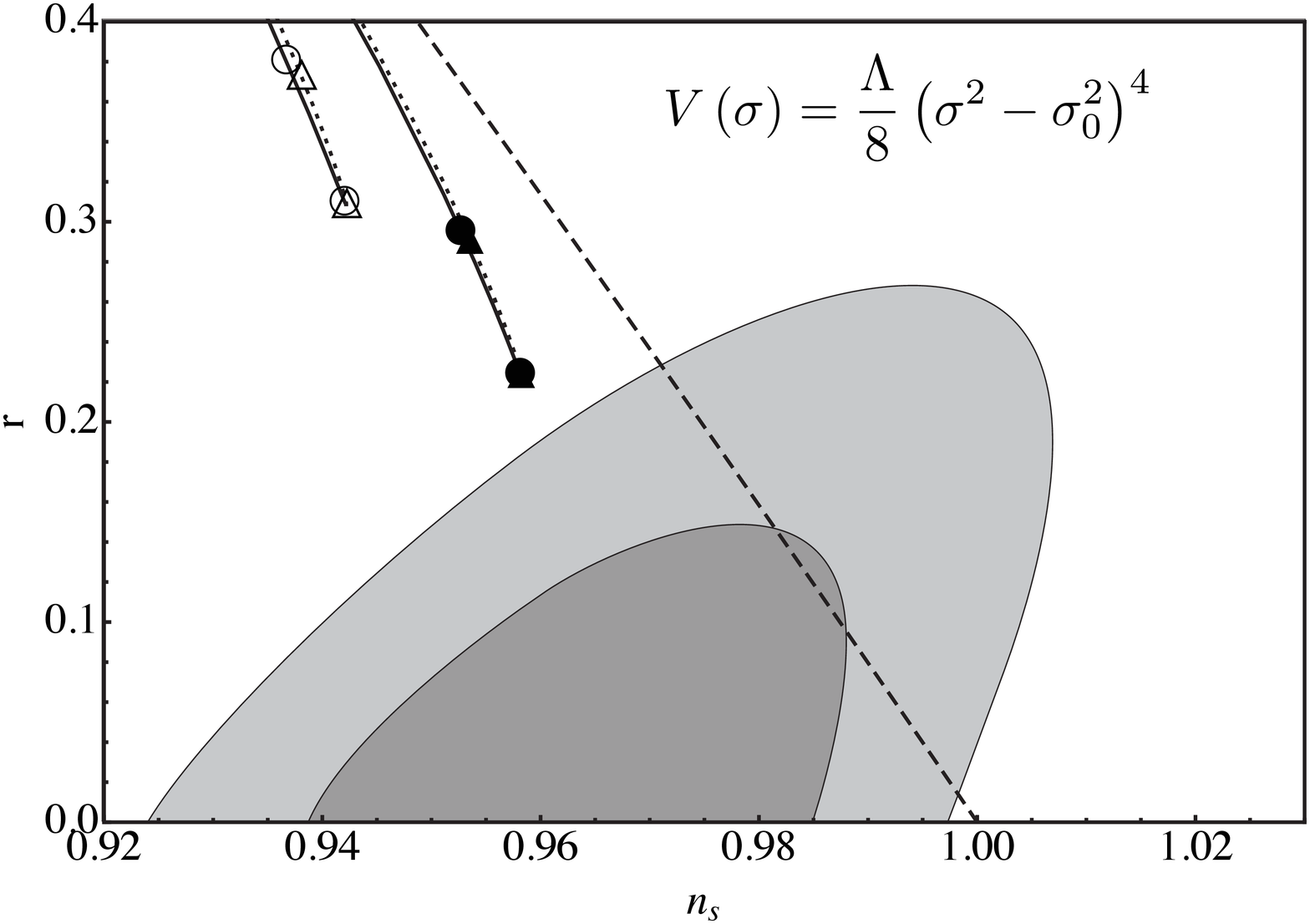, width=8.5 cm}
\epsfig{file=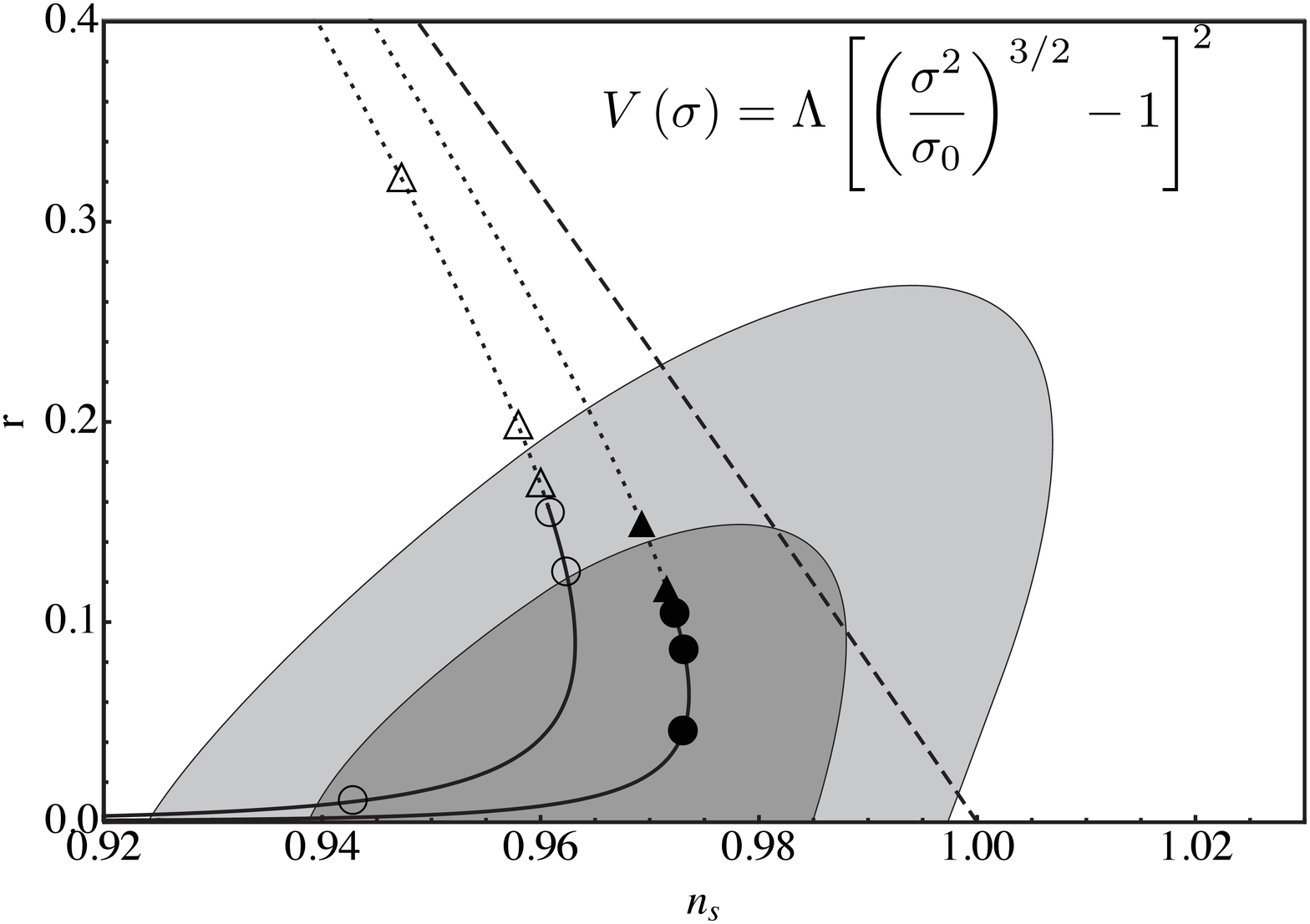, width=8.5 cm}
\epsfig{file=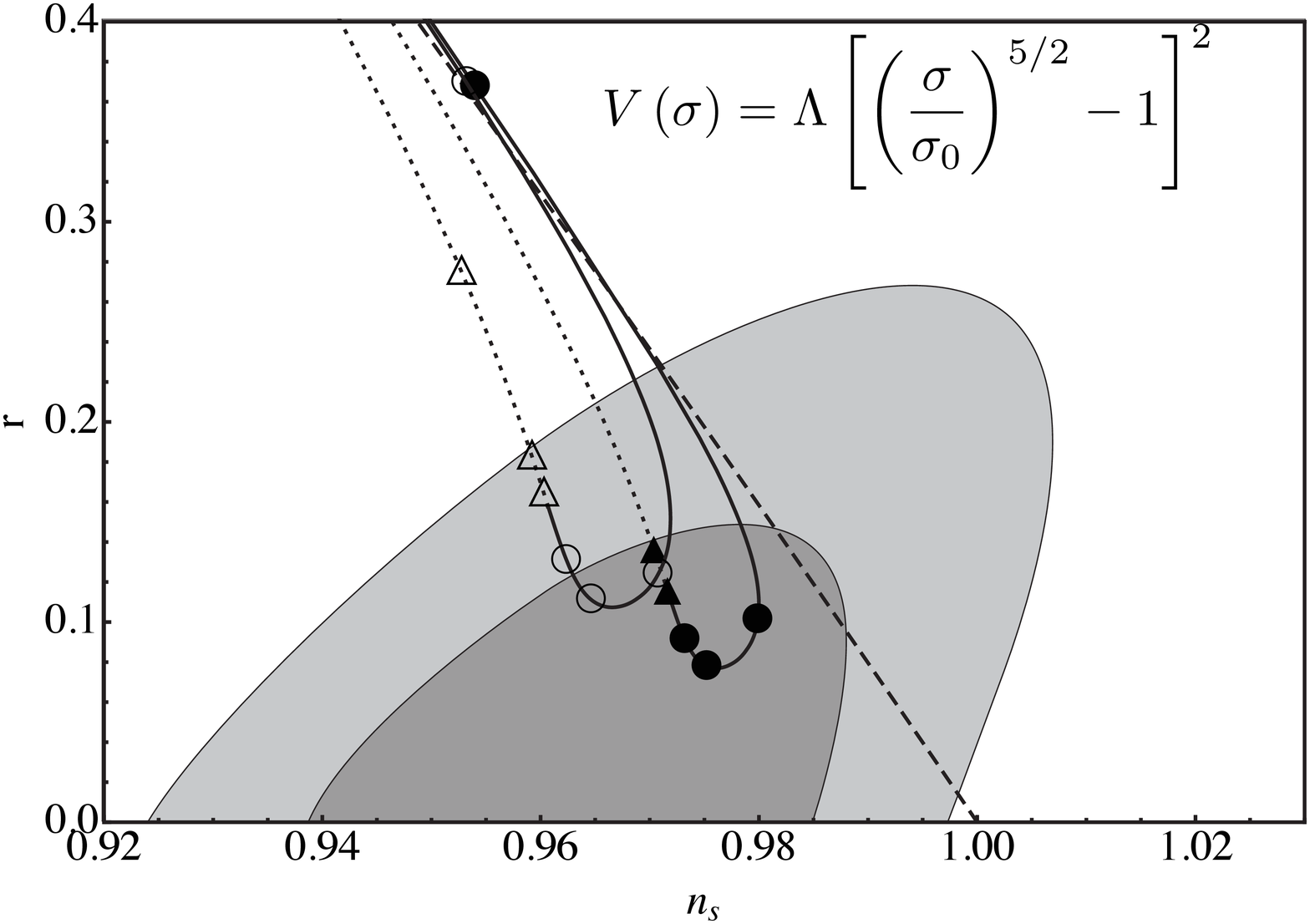, width=8.5 cm}
\caption{{\it The trajectories of the vector $(n_{s},r)$ on 
varying $\gamma$ for potential (\ref{pot1}) with $n=2$ and potential 
(\ref{pot2}) with $n=3/2$ and $n=5/2$ respectively. In the top figure, 
for potential (\ref{pot1}) with $n=2$, the markers represent $L_{\gamma}=-5,-3$ 
both in the SF and in the LF regime. In the middle figure, for potential
(\ref{pot2}) and $n=3/2$, the markers are
$L_{\gamma}=-5,-4,-3$ in the SF regime and $L_{\gamma}=-6,-4,-2$ in the LF regime. In the lowest figure, for potential (\ref{pot2}) and $n=5/2$, the markers are $L_{\gamma}=-5,-4,-3$ in the SF regime and $L_{\gamma}=-3.6,-2.8,-2.0,-1.2$ in the LF regime.} 
\label{rns2}}
\end{figure}

\begin{figure}[t!]
\centering
\epsfig{file=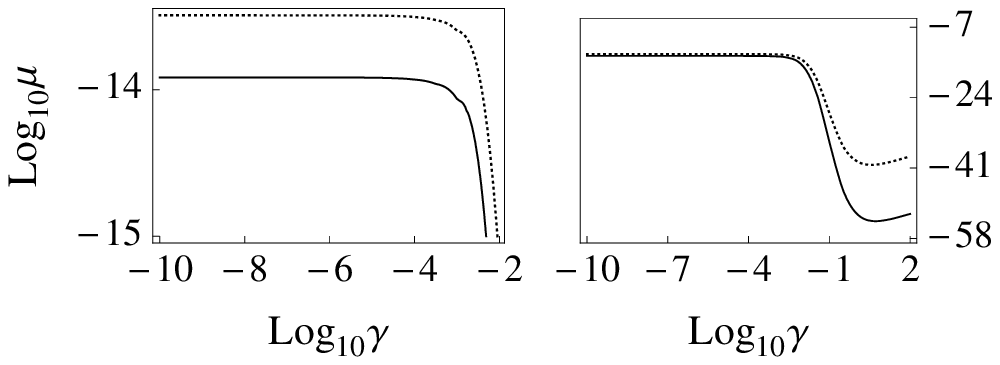, width=8.5 cm}
\epsfig{file=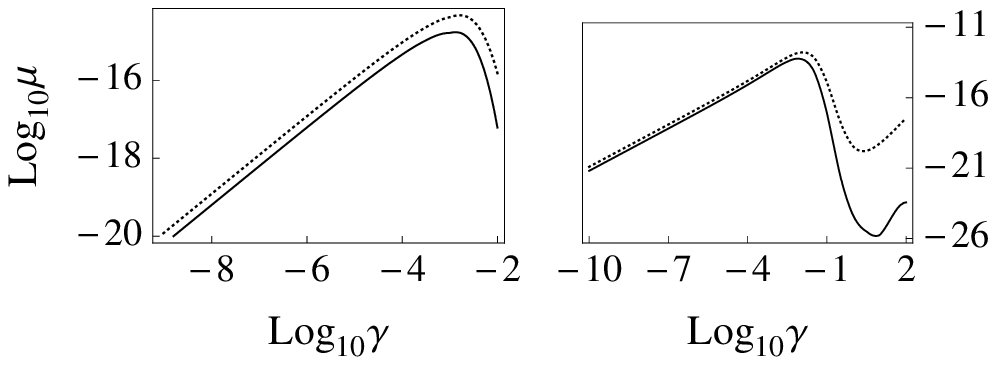, width=8.5 cm}
\epsfig{file=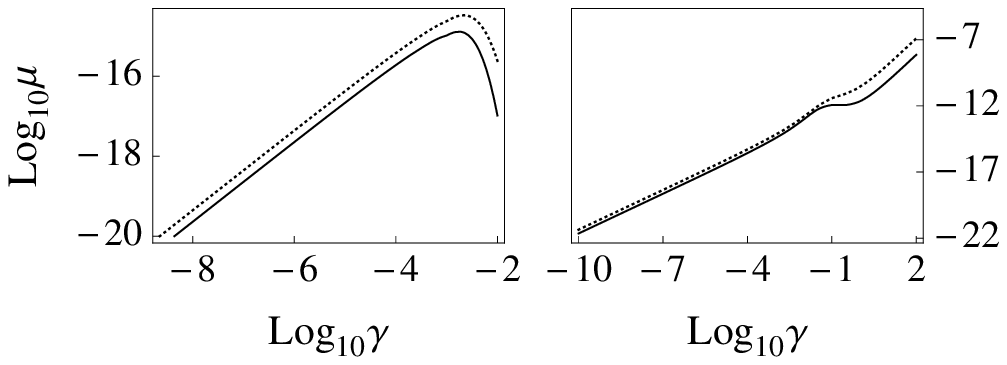, width=8.5 cm}
\caption{{\it Constraints on $\mu$ from the amplitude of the scalar perturbations (\ref{conPR}) for potentials (\ref{pot1}) with $n=2$ and (\ref{pot2}) with $n=3/2$ and $n=5/2$ respectively, where $\mu\equiv\Lambda\,\sigma_{0}^{4}$ for (\ref{pot1}) and $\mu\equiv\Lambda/\sigma_{0}^{4}$ for (\ref{pot2}). The dotted line is for $N_{*}=50$ and the solid line is for $N_{*}=70$ e-folds. The plots on the left refer to the SF case and the ones on the right to the LF.}
\label{PRNI2}}
\end{figure}

On using the above expressions one can investigate the observational constraints 
IG inflation imposes on different potentials for the scalar field. In particular 
we tested the following symmetry breaking potentials which fix the Planck mass after inflation:
a Landau-Ginzburg (LG) potential

\be{LGpot}
V_{LG}\pa{\sigma}=\frac{\mu}{4}\pa{\sigma^{2}-\sigma_{0}^{2}}^{2},
\ee
a Coleman-Weinberg (CW) type potential 
\be{CWpot}
V_{CW}\pa{\sigma}=\frac{\mu}{8}\sigma^{4}\pa{\log\frac{\sigma^{4}}{\sigma_{0}^{4}}-1}+\frac{\mu}{8}\sigma_{0}^{4},
\ee
a cosine potential (CO)
\be{NIpot}
V_{CO}\pa{\sigma}=\Lambda\paq{1+\cos\pa{\pi\frac{\sigma}{\sigma_{0}}}}
\ee
and some generalization of the above potentials of the form
\be{pot1}
V_{1}\pa{\sigma}=\frac{\Lambda}{4n}\pa{\sigma^{2}-\sigma_{0}^{2}}^{2n}
\ee
and
\be{pot2}
V_{2}\pa{\sigma}=\Lambda\paq{\pa{\frac{\sigma}{\sigma_{0}}}^{n}-1}^{2}
\ee
with $n>1$. In particular for our analysis we chose $n=2$ in the case of (\ref{pot1}) and $n=3/2$ and $n=5/2$ in the case of (\ref{pot2}).\\
These potentials share a few interesting features: they have a minimum for $\sigma=\sigma_{0}$ and a relative maximum in $\sigma=0$, $V\pa{\sigma}\ge 0$ and all of them allow both small and large field inflation. We further note that we have included the CO potential even if it agrees to lowest order with the LG potential on expanding around $\sigma_{0}$. Nonetheless it leads to slightly different results because of higher order effects.
On assuming that the pivot mode exits the horizon from 50 to 70 e-folds ($N_{*}$) before inflation ends, one can compare theoretical predictions for $n_{s}-1$ and $r$ with WMAP5+BAO+SN observations \cite{Komatsu:2008hk} for different choices of $\gamma$.\\
In Figures (\ref{rns1},\ref{rns2}) (where the outer border of the lighter grey region represents the 68\% confidence level and the border of the dark grey region the 95\% confidence level) we plotted the trajectories of $(n_{s},r)$ for $N_{*}=50$ (line with empty markers) and $N_{*}=70$ (line with filled markers) as a function of $\gamma$. The markers on the trajectories represents particular choices of $L_{\gamma}\equiv\log{\gamma}$ and each trajectory splits into a dotted line, for the SF regime, and a solid line for the LF regime predictions. Note that the transition between the two regimes simply occurs in the $\gamma\rightarrow 0$ limit.  The dashed line in each figure represents the exact consistency condition (\ref{consrel}) which is $r\,n_{s}=3 r+16\,n_{s}-16$.\\
Potentials (\ref{LGpot},\ref{CWpot}) do not constrain $\gamma$ in the LF regime but they need $L_{\gamma}<-4$ for successful inflation in the SF case. Indeed from Fig. (\ref{rns1}) one can see that the markers for $L_{\gamma}= 3$ lie outside the 68\% confidence lever region.  For different choices of the potential we observe that both the SF and the LF regimes impose limits on $\gamma$ (apart from the potential in (\ref{pot1} with $n=2$) which is not compatible with observations independently of $\gamma$). In particular for $50\le N_{*}\le 70$, $L_{\gamma}\lesssim-4$ is needed in the SF case and $L_{\gamma}\lesssim-2$ in the LF case.\\
One can estimate the coordinate of some relevant points of the trajectories in Figs. (\ref{rns1},\ref{rns2}) as functions of $N_{*}$. On observing than the dynamics of $\sigma$ during inflation is strongly dependent on the value of $\gamma$ (\ref{delta1}, \ref{delta2}) then, in the $\gamma\ll 1$ limit, one can safely assume that $\left|\sigma_{*}-\sigma_{0}\right|/\sigma_{0}\ll1$. In order to relate $\sigma_{*}$ and $N_{*}$ one needs to integrate the SR condition (\ref{delta1}) by parts:
\be{byparts}
\int_{\sigma_{*}}^{\sigma_{0}}\frac{1+\gamma\pa{n(\sigma)+2}}{\gamma\sigma\pa{4-n(\sigma)}}\;\rd\sigma=\int_{0}^{N_{*}}\rd N=N_{*}
\ee
where $n(\sigma)\equiv \rd\log V(\sigma)/\rd \log\sigma$ and we approximate $\sigma_{end}$ (the value of $\sigma$ when inflation ends) with $\sigma_{0}$ since $\sigma_{\rm end}=\sigma_{0}(1\pm\mathcal{O}(\gamma))$. In general, the integral on the l.h.s of Eq.\ (\ref{byparts}) cannot be solved exactly for any potential $V(\sigma)$, and even when it is possible, one further needs to invert the result in order to obtain $\sigma_{*}=\sigma(N_{*})$ which is often a difficult task. For $\gamma\ll 1$, however, simplifications occur and a double series expansion in $\gamma$ and $\pa{\sigma_{*}-\sigma_{0}}/\sigma_{0}$ leads to good predictions. In particular one finds $\sigma\pa{N_{*}}\simeq \sigma_{0}\pa{1-2\sqrt{\gamma N_{*}}}$, $n_{s}\simeq 1-2/N_{*}$ and $r\simeq 8/N_{*}$ for the potentials (\ref{LGpot},\ref{CWpot}) and for (\ref{pot2}) with $n=3/2$ and $n=5/2$, in agreement with the numerical result \footnote{This result differs from that presented previously \cite{CFTV}. The reason is that the initial position of the field for SF inflation depends on the choice of $\gamma$. Indeed for sufficiently small $\gamma$ a small shift of the inflaton from the minimum is sufficient to guarantee the necessary inflation, therefore we have expanded about the minimum.}. Potential (\ref{pot1}) with $n=2$ is quite different and leads to $\sigma\pa{N_{*}}\simeq \sigma_{0}\pa{1-2\sqrt{2\gamma N_{*}}}$, $n_{s}\simeq 1-3/N_{*}$ and $r\simeq 16/N_{*}$ for $\gamma\ll 1$. Thus, depending on the shape of the potential, for small $\gamma$, predictions can be incompatible with observations for a $50-70$ e-fold inflation.\\
For $\gamma$ large inflation takes place in the LF regime and one can still invert Eq.\ (\ref{byparts}) on assuming $\sigma_{*}\gg\sigma_{0}$ and series expanding $\gamma^{-1}$ and $\pa{\sigma_{*}-\sigma_{0}}/\sigma_{*}$ (around zero). One obtains $n_{s}\simeq 1-2/N_{*}$ and $r\simeq 12/N_{*}^{2}$ for potential (\ref{LGpot}) and $n_{s}\simeq 1-1.5/N_{*}$ and $r\simeq 4/N_{*}$ for potential (\ref{CWpot}). In both cases predictions agree with observations but the results are quite different.\\
Furthermore in the Figs. (\ref{PRNI},\ref{PRNI2}) we have plotted the observational constraints on the potentials coming from the amplitude of the scalar perturbations 
\be{conPR}
P_\mathcal{R}(k^*)= (2.445\pm0.096)\times 10^{-9}
\ee
derived from WMAP5 + BAO + SN \cite{Komatsu:2008hk}. The constraints on $\mu$, where $\mu\equiv \Lambda/\sigma_{0}^{4}$ in the case of (\ref{NIpot},\ref{pot2}) and $\mu=\Lambda\,\sigma_{0}^{4}$ in the case of (\ref{pot1}), are plotted for varying $L_{\gamma}$ and assuming $N_{*}=50$ (dotted lines) and $N_{*}=70$ (solid lines).

%%%%%%%%%%%%%%%%%

%%%%%%%%%%%%%%%%
%\section{Reheating}
%Inflation ends when the homogeneous component of scalar field is close enough 
%to the minimum of the potential. In such a regime SR conditions break down 
%and the fields begins to coherently oscillate around the minimum of the potential 
%with an amplitude which decreases in time due to the friction terms in (\ref{KGEQ}). 
%In realistic models of inflation, it is generally assumed that during this phase the 
%inflaton decays into ordinary matter.
%%%%%%%%%%%%%%%

\section{Coherent Oscillations}
In general, in order to avoid never-ending inflation, it is necessary the inflationary potential has a minimum in which case the accelerated epoch is followed by a stage in which the inflaton oscillates about it.
In realistic models of inflation it is usually assumed that 
during this phase the
inflaton decays into ordinary matter.
This regime of coherent oscillations was studied in the context of 
EG in Ref.\ \cite{turner} through the use of a time-averaging procedure which relied on the fact that the frequency of oscillation of the scalar field is much larger than the Hubble parameter.\\ 
%In such an approach coherent field oscillations behave like a fluid and one does not need explicit solutions for scalar field.\\
We have studied the coherent oscillations of the scalar field in the context of IG finding the following solutions \cite{CFTV}:
\ba 
\sigma(t) &=&\sigma_0 + \frac{2}{t} \sqrt{\frac{\gamma}{3\mu}}
\sin \left( \omega_{0} t \right) +
\mathcal{O}\left(\frac{1}{t^2}\right) \\
H (t) &\simeq& \frac{2}{3t}\left[1-\sqrt{\frac{6\gamma}{1+6\gamma}}
\cos \left( \omega_{0} t \right) \right] 
+ \mathcal{O}\left(\frac{1}{t^2}\right)
\label{coherent}
\ea 
for a potential which can be approximated 
\be{quadapprox}
V (\sigma) \simeq  \frac{m^{2}}{2} (\sigma -\sigma_0)^2
\ee
around the minimum. The frequency of the oscillations is given by:
\be{frequency}
\omega_{0} = \sqrt{\frac{m^2}{1+6\gamma}}\,.
\ee
\subsection{MSA Method}
The method we employ relies on the assumption that, during the coherent oscillations phase, two 
different time scales enter the evolution of the field: 
the frequency of the oscillations around the minimum and 
the damping rate of their amplitude, the former being much bigger than the latter. 
This assumption will be verified at the end of the calculations.\\
On using Eqs.\ (\ref{FrEQ},\ref{KGEQ}) one can write the 
second order equation which governs the homogeneous evolution 
of the scalar field in the absence of matter. For a generic potential $V\pa{\sigma}$ one finds
\begin{eqnarray}\label{SFdyn}
\!\!\!\!\!\ddot \sigma &+&\frac{1}{1+6\gamma}\left(\frac{\rd V(\sigma)}{\rd \sigma}-4\frac{V(\sigma)}{\sigma}\right)-2\frac{\dot\sigma^{2}}{\sigma}\nonumber\\
&+&\frac{\dot \sigma}{\sigma}\sqrt{\frac{3}{2\gamma}\left[2\,V(\sigma)+\left(1+6\gamma\right)\dot\sigma^{2}\right]}=0.
\end{eqnarray}
If the potential has a minimum for $\sigma=\sigma_{0}$ then the equation governing the dynamics of $\ds=\sigma-\sigma_{0}$ can be expanded around it up to second order in $\ds/\sigma_{0}$ obtaining
%\begin{eqnarray}\label{SFdynser}
%\ddot\ds&+&\frac{2\mu\sigma_{0}^{2}}{1+6\gamma}\ds-\frac{\mu\sigma_{0}}{1+6\gamma}\ds^{2}-2\frac{\dot\ds^{2}}{\sigma_{0}}\nonumber\\
%&+&\frac{\dot\ds}{\sigma_{0}}\sqrt{\frac{3}{2\gamma}\left[2\mu\sigma_{0}^{2}\ds^{2}+\left(1+6\gamma\right)\dot\ds^{2}\right]}.
%\end{eqnarray}
\begin{eqnarray}\label{SFdynser2}
\ddot{\delta\sigma}& + &\omega_0^2 \delta\sigma + \frac{\bar n-4m^2/\sigma_0}{2(1+6\gamma)} \delta\sigma^2 - 2 \frac{\dot{\delta\sigma}^2}{\sigma_0}\nonumber\\
&+& \frac{\dot{\delta\sigma}}{\sigma_0}\sqrt{\frac{3(1+6\gamma)}{2\gamma}\left[ \omega_0^2\,\delta\sigma^2 + \dot{\delta\sigma}^2\right]}=0
\end{eqnarray}
and for small displacements around the minimum of the potential one has 
\begin{eqnarray}\label{parpot}
V\pa{\sigma}&\equiv&\frac{m^{2}}{2}\ds^{2}+\mathcal{O}\pa{\ds^{3}}\,,\\ 
\frac{\rd V\pa{\sigma}}{\rd \sigma}&\equiv& m^{2}\ds+\frac{\bar n}{2}\ds^{2}+\mathcal{O}\pa{\ds^{3}}
\end{eqnarray}
where $\paq{\bar n}=\paq{m}$.
The effect of the friction terms in ({\ref{KGEQ}}) ensures that $\ds/\sigma_{0}\ll 1$ for some time $\bar t$ and consequently the second order equation is a good approximation for $t>\bar t$. On examining Eq.\ (\ref{SFdynser2}) we observe that second order contributions are responsible for the ``slow'' time scale evolution while the ``fast'' dynamics is that of an harmonic oscillator with frequency $\omega_{0}$. On explicitly taking into account the two time scales, we look for a solution in the form
\be{msexp}
\ds=\ds(t,\tau)=\ds_{0}(t,\tau)+\epsilon\,\ds_{1}(t,\tau)+\mathcal{O}\pa{\epsilon^{2}}
\ee
with $\tau\equiv \epsilon \,t$. On replacing $\ds/\sigma_{0}\rightarrow \epsilon \,\ds/\sigma_{0}$ in (\ref{SFdynser2}) one is finally led to the set of partial differential equations for $\ds_{0}$ and $\ds_{1}$:
\begin{eqnarray}
\!\!\!\!\!\!\!\!\ddot {\ds_{0}}+\omega_{0}^{2}\ds_{0}&=&0\label{mso1}\\
\!\!\!\!\!\!\!\!\ddot {\ds_{1}}+\omega_{0}^{2}\ds_{1}&=&-2\frac{\partial^{2}\ds_{0}}{\partial t\,\partial \tau}-\frac{\bar n-4m^2/\sigma_0}{2(1+6\gamma)} \delta\sigma_0^2+2\frac{\dot{\ds_{0}}^{2}}{\sigma_{0}}\nonumber\\
&&-\frac{\dot{\ds_{0}}}{\sigma_{0}}\sqrt{\frac{3(1+6\gamma)}{2\gamma}\left[\omega_{0}^{2}\,\ds_{0}^{2}+\dot{\ds_{0}}^{2}\right]}\label{mso2}.
\end{eqnarray}
The standard MSA method consists in writing the general solution for (\ref{mso1}) as
\be{solo1}
\ds_{0}=A^{*}\pa{\tau}{\rm e}^{i\,\omega_{0}\,t}+c.c.
\ee
and determining $A\pa{\tau}$ by requiring the cancellation of secular terms in the next to leading order equation (\ref{mso2}) [see \cite{Bender} for details]. From this procedure one obtains the following differential equation for $A\pa{\tau}$
\be{eqA}
\frac{\rd A}{\rd\tau}+\sqrt{\frac{3\pa{1+6\gamma}}{2\gamma}}\frac{\omega_{0}}{\sigma_{0}}\left|A\right|A=0
\ee
and on setting $A\pa{\tau}=R\pa{\tau}{\rm e}^{i\,\theta\pa{\tau}}$ one finds $\theta\pa{\tau}=\theta_{0}$ and
\be{solR}
R\pa{\tau}=\frac{R_{0}}{1+\frac{R_{0}}{\sigma_{0}}\sqrt{\frac{3\pa{1+6\gamma}}{2\gamma}}\,\omega_{0}\,\tau}=\sigma_{0}\frac{\,f\,r}{1+r\,\omega_{0}\,t}
\ee
where 
\be{defMSA}
f\equiv\sqrt{\frac{2\gamma}{3\pa{1+6\gamma}}}\,,\quad r\equiv \frac{\Omega_{0}}{\omega_{0}}=\frac{R_{0}}{f\,\sigma_{0}}
\ee
and $\Omega_{0}=R_{0}\sqrt{3\mu/\gamma}$ is the inverse of the ``slow'' time scale. Let us note that the ratio $r$ between the two time scales depends on the ratio $R_{0}/\sigma_{0}$ and on $f$, where $f$ is a function of $\gamma$ having values between $0$ and $\sqrt{1/3}$. On setting $\sigma_{0}=\M/\sqrt{\gamma}$ one finds
\be{rsmall}
r=\sqrt{\frac{1+6\gamma}{2}}\frac{R_{0}}{\M}\ll 1\Longrightarrow \frac{R_{0}}{\M}\ll\sqrt{\frac{2}{1+6\gamma}}
\ee
which is the condition for the MSA method to well approximate the dynamics. 
The general solution for (\ref{SFdynser2}) turns out to be 
\be{solmsa}
\ds=\sigma_{0}\frac{2\,f\,r}{1+r\,\omega_{0}\,t}\cos\pa{\omega_{0}\,t+\theta_{0}}.
\ee
and the constants $R_{0}$ and $\theta_{0}$ are related to the initial conditions of the second order equation (\ref{SFdynser2}). \\
The MSA method is generally applied to first order, nonetheless we found numerically that our results were significantly improved once the second order contributions on the r.h.s. of Eq.\ (\ref{mso2}) were taken into account. These terms slowly vary in time and give a second order shift to the centre of the oscillations which has to be added to the r.h.s. of (\ref{solmsa}) 
%The MSA method is generally applied to first order, as we have, since, in general, second order calculations do not improve the first order results. Nonetheless we found a second order refinement of our results by considering the non-oscillating contributions in the r.h.s. of Eq.\ (\ref{mso2}). These terms slowly vary in time and give a second order shift to the centre of the oscillations which is to be added to the r.h.s. of (\ref{solmsa}) 
%\be{shiftLG}
%\Delta_{LG}=5\sigma_{0}\frac{f^{2}r^{2}}{\paq{1+r\,\omega_{0}\,t}^{2}}
%\ee
\be{shiftGEN}
\Delta_{\bar n}=|A|^2 \left(\frac{8}{\sigma_0}-\frac{\bar n}{m^2} \right)
\ee
and plays a crucial role in the calculations of the next sections. In particular one has $\bar n_{LG}=6\,\mu \,\sigma_{0}$ and $\bar n_{CW}=10\,\mu\,\sigma_{0}$ and $m^{2}=2\mu\sigma_{0}^{2}$ for both LG and CW potentials.\\
%%%%%%%%%%%%%%%%%%%%
\begin{figure}[t!]
\centering
\epsfig{file=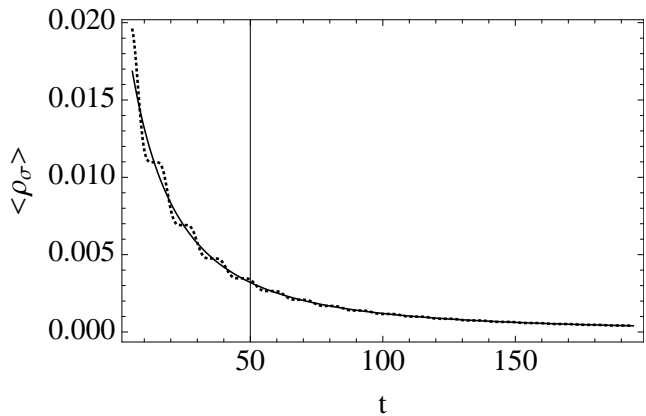, width=8.5 cm}
\caption{{\it An example of the average value of the energy density of the scalar field, as given in 
(\ref{enpres}), for (\ref{LGpot}), $\gamma=10$, $\mu=10$ calculated exactly (dotted line) and with MSA in (\ref{H2LGav}) (solid line). The plot shows good agreement between exact and approximate solutions.} 
\label{LGh2MED}}
\end{figure}
\subsection{Equation of State}
\begin{figure}[t!]
\centering
\epsfig{file=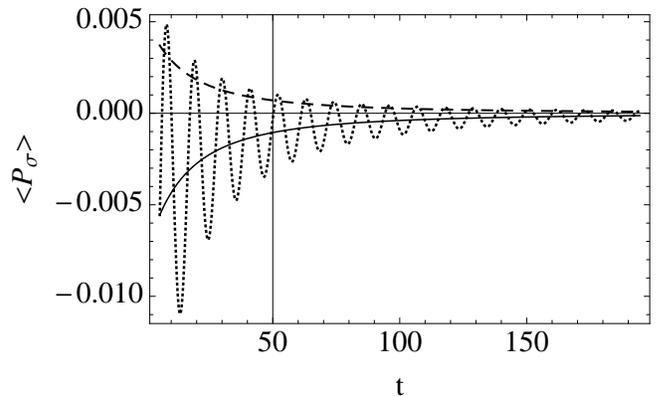, width=8.5 cm}
\caption{{\it In the above plot examples are given of the average pressure versus time for the potential (\ref{LGpot}) with $\gamma=10$ and $\mu=10$. The dotted line corresponds to both the average value of the pressure of the scalar field as defined in (\ref{enpres}) calculated exactly (numerically) and the plot of (\ref{dHLGav}), since both coincide within the resolution of the figure. The solid line is the average of (\ref{dHLGav}) and the dashed line is the average of the pressure calculated without considering the correction term (\ref{shiftGEN}). The plot shows good agreement between exact and approximate values.}
%\caption{{\it In the above plots examples are given of the average pressure versus time. In the figure on top the average value of the pressure of the scalar field as defined in (\ref{enpres}) for potential (\ref{LGpot}) with $\gamma=10$, $\mu=10$ is calculated exactly (dotted line) and is compared with the average of (\ref{dHLGav}) (solid line). In the lower figure the dotted line is a plot of (\ref{dHLGav}), the solid line is its average and the dashed line is the average of the pressure calculated without considering the correction term (\ref{shiftGEN}). The plots show good agreement between exact and approximate values.} 
\label{LGIMP}}
\end{figure}

A first application of the above results 
is the calculation of the averaged equation of state of the scalar field during the oscillatory regime. It is known (and can be shown with the same technique we are employing) that the averaged massive oscillations of a scalar field in EG are equivalent to a fluid with null equation of state in agreement with \cite{turner}. On the other hand, in IG, for the potentials we are considering, oscillations around the minimum are still massive but the results are somewhat different.\\
The definition of energy density and pressure for a scalar field in IG can be given \cite{FTV} further
in the EG form:
\be{enpres}
\rho_{\sigma}\equiv3\gamma\sigma_{0}^{2}H^{2},\quad P_{\sigma}\equiv-2\gamma\sigma_{0}^{2}\pa{\dot H+\frac{3}{2}H^{2}}.
\ee
With the above definitions continuity equation is satisfied. At the end of inflation the expressions on the r.h.s. of (\ref{enpres}) are oscillating functions and one may define the averaged quantities $\av{\rho_{\sigma}}$ and $\av{P_{\sigma}}$ where 
\be{defav}
\av{A(t)}=\frac{1}{T}\int_{t-T/2}^{t+T/2}A(t')\,\rd t'
\ee
and $T=2\pi/\omega_{0}$ is the period of oscillation. Further one may define the averaged equation of state by
\be{defEOS}
w\equiv\frac{\av{P_{\sigma}}}{\av{\rho_{\sigma}}}.
\ee
On averaging the expressions (\ref{enpres}), using the solution for $\ds+\Delta_{\bar n}$ and keeping contributions up to the second order in $R(t)/\sigma_{0}\ll1$ one finds
\be{H2LGav}
\av{\rho_{\sigma}}=(1+9\gamma)\av{\dot\ds^{2}}\ee
and
\be{dHLGav}
\av{P_{\sigma}}=-\paq{3\gamma+\sqrt{6\gamma\pa{1+6\gamma}}\sin\omega_{0}\,t}\av{\dot\ds^{2}}
\ee
where 
\be{avsigmadotsq}
\av{\dot\ds^{2}}=4\,\omega_{0}\,\sigma_{0}\frac{f^{2}r^{2}}{\pa{1+r\,\omega_{0}\,t}^{2}}
\ee
leading to
\be{eqofstate}
\av{w}=-\frac{3\gamma}{1+9\gamma}.
\ee
%In Fig.\ (\ref{LGh2MED}) we plot the average $\av{\rho_{\sigma}}$ calculated numerically (dotted line) and analytically as in Eq. (\ref{H2LGav}) for the potential (\ref{LGpot}) with $\gamma=10$ and $\mu=10$. The plot shows that multiple scale predictions agree well with the corresponding exact quantities. Good agreement between the analytical predictions in Eq. (\ref{dHLGav}) and the numerical solutions is also evident in Figs.\ (\ref{LGIMP}) for potential (\ref{LGpot}) with $\gamma=10$ and $\mu=10$. In the upper plot the dotted line represents the numerical (exact) expression and the solid line is the average of (\ref{dHLGav}); in the lower plot we exhibit the expression (\ref{dHLGav}), its average and the average of $\av{P_{\sigma}}$ without adding the improvement (\ref{shiftGEN}) to (\ref{solmsa}).\\
In Fig.\ (\ref{LGh2MED}) we plot the average $\av{\rho_{\sigma}}$ calculated numerically (dotted line) and analytically as in Eq. (\ref{H2LGav}) for the potential (\ref{LGpot}) with $\gamma=10$ and $\mu=10$. The plot shows that multiple scale predictions agree well with the corresponding exact quantities. Good agreement between the analytical predictions of Eq. (\ref{dHLGav}) and the numerical solutions is also evident in Fig.\ (\ref{LGIMP}) for the potential (\ref{LGpot}) with $\gamma=10$ and $\mu=10$. In this plot the dotted line represents both the numerical (exact) expression and its analytical approximation (\ref{dHLGav}), since both coincide within the resolution of the figure, the solid and the dashed lines are respectively the average of $\av{P_{\sigma}}$ with and without adding the improvement (\ref{shiftGEN}) to (\ref{solmsa}).\\
In contrast with EG, in IG the averaged equation of state defined by (\ref{defEOS}) does not lead to the correct averaged expansion, as can be directly verified on using (\ref{coherent}). The averaged Hubble parameter is simply $\av{H(t)}\simeq 2/(3t)$, namely that of a matter dominated Universe, independently of the equation of state (\ref{eqofstate}). This difference is a consequence of the fact that in EG one can verify that $\av{H_{(EG)}}\simeq \sqrt{\av{H_{(EG)}^{2}}}$ whereas in IG such a relation does not hold.
%%%%%%%%%%%%%%%%%%%%%%%%

\section{Reheating}
On assuming that the inflaton field is coupled 
to ordinary matter, then at the end of 
inflation such a scalar field can decay into 
lighter particles losing its energy density 
and eventually reheating the Universe. 
Such a process has been traditionally described through a phenomenological 
decay width $\Gamma$ which enters the equation for the scalar field as an additional friction term:
\be{modkgreh}
\ddot\sigma+\pa{3H+\Gamma}\dot \sigma+V_{,\sigma}-
6\gamma\pa{H^{2}+\frac{\ddot a}{a}}\sigma=0.
\ee 
On requiring that the newly added friction term does not affect the Einstein equations
(the Bianchi identities must still hold) one observes that the continuity equation for radiation should also be modified as:
\be{contRAD}
\dot \rho_{R}=-3H\pa{\rho_{R}+P_{R}}+\Gamma\dot\sigma^{2}.
\ee

The system of equations (\ref{modkgreh},\ref{contRAD}) can be solved numerically 
but analytical solutions are needed to estimate the reheating temperature of the relativistic 
particle fluid.\\
In order to simulate the energy transfer from the inflaton to radiation and estimate 
the reheating temperature, we need the definition of the energy density and pressure 
of the scalar field. In the presence of relativistic matter the expressions for the energy density and 
pressure of the scalar field, Eqs. (\ref{enpres}), must be modified as
\begin{eqnarray}
\rho_{\sigma}&=&3\gamma\sigma_{0}^{2}H^{2}-\rho_{R}\label{energyRH}\\
P_{\sigma}&=&-2\gamma\sigma_{0}^{2}\pa{\dot H+\frac{3}{2}H^{2}}-\rho_{R}-P_{R};\label{pressureRH}
\end{eqnarray}
with these definitions the continuity equation for $\rho_{\sigma}$, $P_{\sigma}$ is still automatically conserved and the introduction of the decay-rate $\Gamma$ leads to
\be{contSIGMA}
\dot \rho_{\sigma}=-3H\pa{\rho_{\sigma}+P_{\sigma}}-\Gamma \dot \sigma^{2}.
\ee
Let us note that Eqs. (\ref{contRAD}) and (\ref{contSIGMA}) are formally the same as in EG, but they lead to different predictions.\\

The reheating temperature can be estimated on considering the evolution of $\rho_{R}$ and $\av{\rho_{\sigma}}$ as $T_{\rm RH}\sim \rho_{R}^{1/4}$ when $\rho_{R}\gtrsim \av{\rho_{\sigma}}$. In EG one can naively observe that this happens when $3H\sim\Gamma$ and one finds the well-known estimate $T_{RH}^{(EG)}\sim\sqrt{\Gamma \M}$.\\

The continuity equations for $\rho_{R}$ and $\av{\rho_{\sigma}}$ can be estimated by the MSA method. Starting from the modified equation for $\sigma$
\begin{widetext}
\be{SFdynR}
\!\!\!\!\!\ddot \sigma +\frac{1}{1+6\gamma}\left(\frac{\rd V(\sigma)}{\rd \sigma}-4\frac{V(\sigma)}{\sigma}\right)-2\frac{\dot\sigma^{2}}{\sigma}\nonumber\\
+\frac{\dot \sigma}{\sigma}\sqrt{\frac{3}{2\gamma}\left[2\,V(\sigma)+\left(1+6\gamma\right)\dot\sigma^{2}\right)+2\,\rho_R]}+\frac{\Gamma}{1+6\gamma}\,\dot\sigma=0.
\ee
\end{widetext}
one can expand to second order in $\ds\equiv\sigma-\sigma_{0}$ and consider $\rho_{R}\sim V\pa{\sigma}\sim\dot\ds^{2}$ and $\Gamma\sim H\sim \dot\ds/\sigma_{0}$. These assumptions are reasonable for the regime we are considering \footnote{$\rho_{R}$ is negligible at the beginning of the reheating era and increases becoming comparable with $\rho_{\sigma}$.}, nonetheless they will be checked numerically. On expanding one finds
\begin{eqnarray}\label{SFdynserRH}
\ddot{\delta\sigma}& + &\omega_0^2 \delta\sigma + \frac{\bar n-4m^2/\sigma_0}{2(1+6\gamma)} \delta\sigma^2 - 2 \frac{\dot{\delta\sigma}^2}{\sigma_0}+\frac{\Gamma}{1+6\gamma}\dot{\delta\sigma}
+ \frac{\dot{\delta\sigma}}{\sigma_0}\nonumber\\
&\times&\sqrt{\frac{3(1+6\gamma)}{2\gamma}\left[ \omega_0^2\,\delta\sigma^2 + \dot{\delta\sigma}^2+\frac{2\,\rho_R}{1+6\gamma}\right]}=0
\end{eqnarray}
%\begin{eqnarray}\label{SFdynserRH}
%\!\!\!\!\!\!\!\!\!\!\!\ddot\ds&+&\omega_{0}^{2}\,\ds-\frac{\omega_{0}^{2}}{2}\,\frac{\ds^{2}}{\sigma_{0}}-2\frac{\dot\ds^{2}}{\sigma_{0}}+\frac{\Gamma}{1+6\gamma}\,\dot\ds+\frac{\dot\ds}{\sigma_{0}}\nonumber\\
%&\times&\sqrt{\frac{3(1+6\gamma)}{2\gamma}\left[\omega_{0}^{2}\,\ds^{2}+\dot\ds^{2}+\frac{2\,\rho_{R}}{1+6\gamma}\right]}=0.
%\end{eqnarray}
If the ``fast'' time evolution is still determined by the leading order terms $\ddot\ds_{0}+\omega_{0}^{2}\,\ds_{0}=0$ then, on setting $\ds_{0}=A\pa{\tau}\,y(t)$ and requiring the cancellation of singularities, one obtains the differential equation
\be{seccanc}
2\dot A+\frac{\Gamma}{1+6\gamma}\, A+3\sqrt{\frac{\rho_{R}+\pa{1+6\gamma}A^{2}\,E_{F}
}{3\gamma\sigma_{0}^{2}}}A=0
\ee
where $E_{F}\equiv\frac{1}{2}\dot y^{2}+\frac{\omega_{0}^{2}}{2}y^{2}$. On then keeping the
second order contributions in the continuity Eq.\ (\ref{contRAD}) and averaging over ``fast'' oscillations of $y$ one then finds
\be{contRAD2}
\dot \rho_{R}-\Gamma \,A \,E_{F}+4\sqrt{\frac{\rho_{R}+\pa{1+6\gamma}A^{2}\,E_{F}
}{3\gamma\sigma_{0}^{2}}}\rho_{R}=0
\ee
where $\av{\dot\ds^{2}}=A^{2}E_{F}$. On noting that
\be{avH}
\sqrt{\frac{\rho_{R}+\pa{1+6\gamma}A^{2}\,E_{F}}{3\gamma\sigma_{0}^{2}}}=\av{H}
\ee
and
\be{avenergyRH}
\av{\rho_{\sigma}}=A^{2}E_{F}\pa{1+9\gamma}
\ee
where $H$ and $\rho_{\sigma}$ are expanded to the second order, one is finally led to the system of averaged equations
\begin{eqnarray}
\dot \rho_{R}&=&-4\av{H} \rho_{R}+\frac{\Gamma}{1+9\gamma}\av{\rho_{\sigma}}\label{contRADav}\\
\av{\dot\rho_{\sigma}}&=&-3\av{H} \av{\rho_{\sigma}}-\frac{\Gamma}{1+6\gamma}\av{\rho_{\sigma}}\label{contSIGMAav}\\
\av{H}&=&\sqrt{\frac{\rho_{R}+\frac{1+6\gamma}{1+9\gamma}\av{\rho_{\sigma}}}{3\gamma\sigma_{0}^{2}}}\label{defavH}
\end{eqnarray}
with $\frac{\rd \av{\rho_{\sigma}}}{\rd t}=\av{\dot\rho_{\sigma}}$. \\
On comparing the above equations with those arising in EG for a massive inflaton we observe that averaged energy transfer is less efficient  because of the factors $(1+9\gamma)^{-1}$ and $(1+6\gamma)^{-1}$ in (\ref{contRADav},\ref{contSIGMAav}).\\
\begin{figure}[t!]
\centering
\epsfig{file=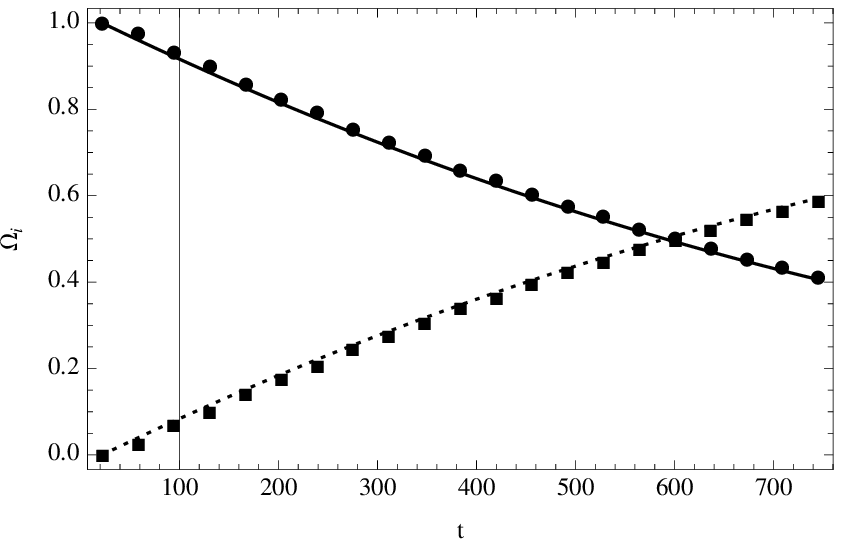, width=8.5 cm}
\epsfig{file=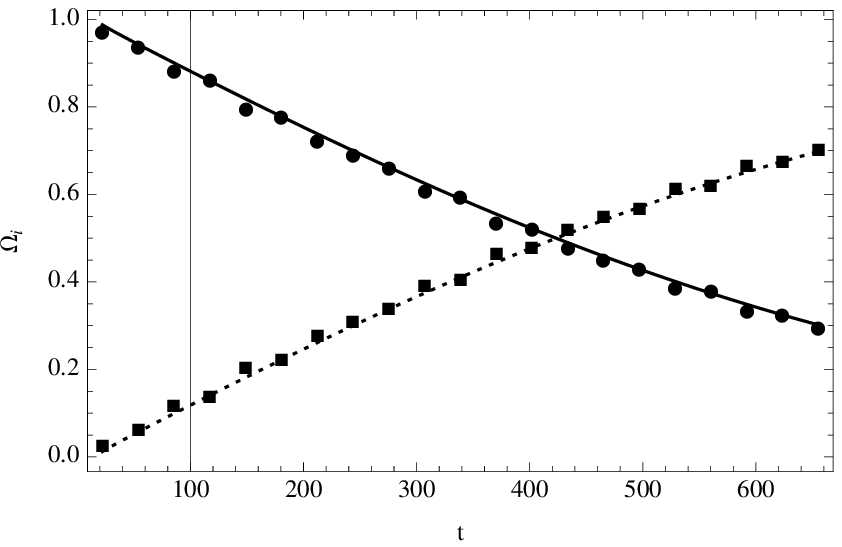, width=8.5 cm}
\caption{{\it In these figures the solid line is for (\ref{omegaS}) and the dotted line is for (\ref{omegaR}) where both are calculated as numerical solutions to Eqs.\ (\ref{contRADav}-\ref{defavH}). Circles and squares represents the same quantities evaluated on solving (\ref{SFdynR},\ref{contRAD}) and then averaging the definition (\ref{energyRH}). The figure on top is for $\Gamma=2\cdot10^{-3}\M$, $\gamma=10^{-2}$ and the lower figure is for $\Gamma=2\cdot 10^{-1}\M$, $\gamma=10$. Both situations  are described well by the approximation scheme used. Time evolution is expressed in $\M^{-1}$ units.} 
\label{LGREH}}
\end{figure}
In Figs. (\ref{LGREH}) we have plotted the comparison of the energy transfer obtained numerically on solving (\ref{modkgreh}) and (\ref{contRAD}) with that calculated by numerically solving the averaged equations (\ref{contRADav}-\ref{defavH}). The continuous line is 
\be{omegaS}
\Omega_{\sigma}\equiv \frac{\av{\rho_{\sigma}}}{\rho_{R}+\av{\rho_\sigma}},
\ee
and the dashed line is 
\be{omegaR}
\Omega_{R}\equiv \frac{\rho_{R}}{\rho_{R}+\av{\rho_\sigma}}.
\ee 
The round dots plot $\Omega_{\sigma}$ and the square dots plot $\Omega_{R}$ where the averages are calculated numerically from the exact numerical solutions by using (\ref{defav}). In particular the plots are obtained on setting $\gamma\sigma_{0}^{2}=\M^{2}$ for $\Gamma=2\cdot10^{-3}\M$, $\gamma=10^{-2}$ (figure on top) and $\Gamma=2\cdot 10^{-1}\M$, $\gamma=10$ (lower figure). In both regimes Eqs.\ (\ref{contRADav}-\ref{defavH}) describe well the exact dynamics and, qualitatively, it appears that particle production is enhanced for $\gamma$ small (when the EG predictions are recovered).\\
%%%%%%%%%%%%%%%
\subsection{The Reheating Temperature}
\begin{figure}[t!]
\centering
\epsfig{file=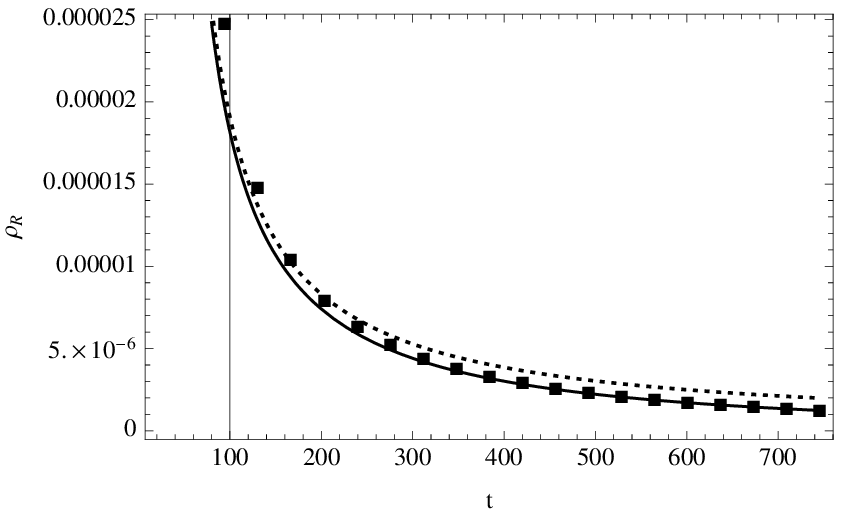, width=8.5 cm}
\epsfig{file=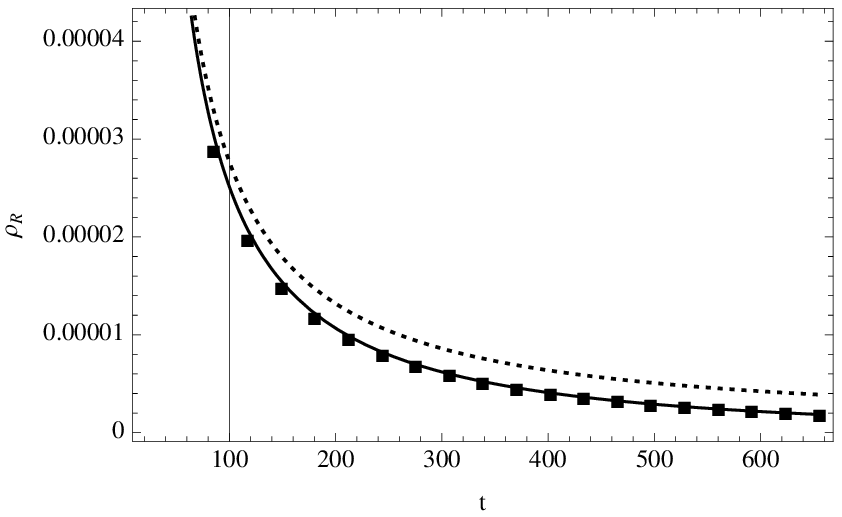, width=8.5 cm}
\caption{{\it In the above figures the continuous line is $\rho_{R}$ (in $\M^{4}$ units) as a function of $t$ (in $\M^{-1}$ units) obtained on numerically solving Eqs.\ (\ref{contRADav}-\ref{defavH}). The squares represent the same quantity evaluated by numerically solving (\ref{SFdynR},\ref{contRAD}) and the dotted line is $\rho_{R}$ calculated from the numerical solution of (\ref{contTRAD},\ref{contTSIGMA}). The figure on top is for $\Gamma=2\cdot10^{-3}\M$, $\gamma=10^{-2}$ and the lower figure is for $\Gamma=2\cdot 10^{-1}\M$, $\gamma=10$. The figures show that assumptions leading to the expression (\ref{solcontRAD}) yield good approximate results independently on the choice of $\gamma$.} 
\label{RHORREH}}
\end{figure}
\begin{figure}[t!]
\centering
\epsfig{file=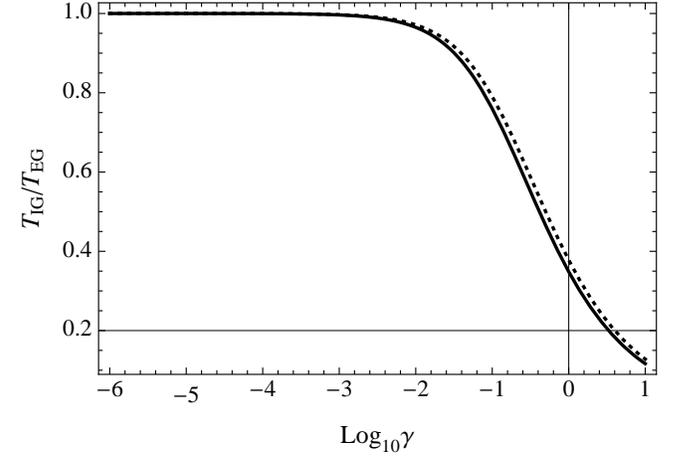, width=8.5 cm}
\caption{{\it The continuous line represents the ratio (\ref{rtemp}) evaluated numerically on using the exact equations for the time of equality $t_{*}$. The dotted line is the analytical estimate of the same temperature ratio obtained from (\ref{rhoR},\ref{rhoREG}) as $T_{\rm reh}^{(IG)}/T_{\rm reh}^{(EG)}=\pa{1+6\gamma}^{-1/2}$.} 
\label{rattemp}}
\end{figure}

We are interested in estimating the energy density of radiation when it starts dominating over the scalar field. At the onset of this stage the averaged Friedmann equation (\ref{defavH}) can be approximated by
\be{avHapprox}
\av{H}\simeq\sqrt{\frac{\frac{1+6\gamma}{1+9\gamma}\av{\rho_{\sigma}}}{3\gamma\sigma_{0}^{2}}}
\ee
and the system of equations (\ref{contRADav},\ref{contSIGMAav},\ref{avHapprox}) can be solved analytically.\\
On defining $\tilde \rho_{\sigma}\equiv \frac{1+6\gamma}{1+9\gamma}\av{\rho_{\sigma}}$ and $\tilde \Gamma=\pa{1+6\gamma}^{-1}\Gamma$ Eqs.\ (\ref{contRADav},\ref{contSIGMAav}) take the form
\begin{eqnarray}
\frac{\rd \rho_{R}}{\rd t}&=&-4\sqrt{\frac{\tilde\rho_{\sigma}}{3\gamma\sigma_{0}^{2}}}\,\rho_{R}+\tilde\Gamma\,\tilde\rho_{\sigma}\label{contTRAD}\\
\frac{\rd \tilde\rho_{\sigma}}{\rd t}&=&-3\sqrt{\frac{\tilde\rho_{\sigma}}{3\gamma\sigma_{0}^{2}}}\,\tilde\rho_{\sigma}-\tilde\Gamma\,\tilde\rho_{\sigma}\label{contTSIGMA}
\end{eqnarray}
which is exactly the form one has in EG for a minimally coupled, massive scalar field. As a consequence of this formal analogy one expects $\tilde \rho_{\sigma}\sim\rho_{R}$ when $\sqrt{3\tilde\rho_{\sigma}}\simeq \Gamma \M/\pa{1+6\gamma}$ or equivalently $t\sim\tilde\Gamma^{-1}$.\\
Eq.\ (\ref{contTSIGMA}) can be solved easily and leads to:
\be{solcontSIGMA}
\av{\rho_{\sigma}}=\frac{\gamma\sigma_{0}^{2}}{3}\frac{1+9\gamma}{\pa{1+6\gamma}^{3}}\frac{\Gamma^{2}}{\paq{C_{0}\exp\pa{\frac{\Gamma}{2\pa{1+6\gamma}}t}-1}^{2}}
\ee
where $C_{0}$ is an integration constant and we assumed that the energy transfer begins at $t=0$. 
On setting
\be{inicond}
C_{0}=1+\frac{\Gamma}{3}\sqrt{\frac{3\gamma\sigma_{0}^{2}}{\av{\rho_{\sigma}}_{0}}\cdot\frac{1+9\gamma}{\pa{1+6\gamma}^{3}}}
\ee
one can explicitly obtain the dependence of $C_{0}$ on the initial condition for the energy density $\av{\rho_{\sigma}}_{0}$.
Further Eq.\ (\ref{contTRAD}) can also be solved and on setting $\rho_{R}\pa{0}=0$ one obtains:
\begin{widetext}
\be{solcontRAD}
\rho_{R}=\frac{\Gamma^{2}\,\gamma\,\sigma_{0}^{2}}{20\pa{1+6\gamma}^{2}}\frac{\left[ \pa{\re^{\frac{\Gamma t}{2\pa{1+6\gamma}}}B_{0}-1}^{5/3}\pa{3\,\re^{\frac{\Gamma t}{2\pa{1+6\gamma}}}B_{0}+5}-b_{0}^{5/3}\re^{\frac{4\Gamma t}{3\pa{1+6\gamma}}}\pa{3\,b_{0}+8}\right]}{\pa{\re^{\frac{\Gamma t}{2\pa{1+6\gamma}}}B_{0}-1}^{8/3}}
\ee
\end{widetext}
where 
\be{defB0}
B_{0}=1+\Gamma\sqrt{\frac{\gamma\,\sigma_{0}^{2}}{3\av{\rho_{\sigma}}_{0}}\cdot\frac{1+9\gamma}{\pa{1+6\gamma}^{2}}}
\ee
and $b_{0}=B_{0}-1$. If one now considers $b_{0}$ and sets $\gamma\, \sigma_{0}^{2}=\M^{2}$ then
\be{b0est}
b_{0}=\frac{\Gamma\,\M}{\av{\rho_{\sigma}}_{0}^{1/2}}\sqrt{\frac{1+9\gamma}{3\pa{1+6\gamma}^{3}}}\simeq\mathcal{O}\pa{\frac{\Gamma}{\av{H}_{0}}}\ll 1
\ee
since it is generally assumed that $\Gamma\ll\av{H_{0}}$, where $\av{H_{0}}$ is the average value of $H$ at the beginning of reheating. The expression (\ref{solcontRAD}) then simplifies and becomes
\be{solRADapp}
\rho_{R}\simeq\frac{\Gamma^{2}\,\gamma\,\sigma_{0}^{2}}{20\pa{1+6\gamma}^{2}}\cdot\frac{3\,\re^{\frac{\Gamma\,t}{2\pa{1+6\gamma}}}+5}{\re^{\frac{\Gamma\,t}{2\pa{1+6\gamma}}}-1};
\ee
and on evaluating it for $t\sim\pa{1+6\gamma}\Gamma^{-1}$ one finally obtains:
\be{rhoR}
\rho_{R}\pa{\pa{1+6\gamma}\Gamma^{-1}}\simeq \frac{3\,\Gamma^{2}\,\M^{2}}{\pa{1+6\gamma}^{2}}.
\ee
On performing the same calculations as above in EG one obtains
\be{rhoREG}
\rho_{R}^{\pa{EG}}\pa{\Gamma^{-1}}\simeq 3\,\Gamma^{2}\,\M^{2}
\ee
which is the standard result. In Figure (\ref{rattemp}) we plot the ratio between the reheating temperatures in IG and in EG on varying $\gamma$. Such a ratio is obtained on evaluating 
\be{rtemp}
\frac{T_{\rm reh}^{(IG)}}{T_{\rm reh}^{(EG)}}=\paq{\frac{\rho_{R}^{(IG)}\pa{t_{*,IG}}}{\rho_{R}^{(EG)}\pa{t_{*,EG}}}}^{1/4},
\ee
where $\rho_{R}^{(IG)}$ and $\rho_{R}^{(EG)}$ are the numerical solutions of the exact equations and $t_{*}$ is the time for which $\rho_{R}\pa{t_{*}}=\rho_{\sigma}\pa{t_{*}}$, and the comparison is made with the analytical prediction obtained from (\ref{rhoR},\ref{rhoREG}). Again we observe that the correct behavior is reproduced by our analytical estimates.
%%%%%%%%%%%%%%%%%%%%%

\section{Preheating}

The phenomenological study of reheating described 
in the previous section was improved by the application 
of the theory of parametric resonance in an expanding universe to 
the fields coupled to the inflaton \cite{TB,KLS}. The first stage of the 
decay of the inflaton - dubbed {\em preheating} - 
can be rather efficient and leads to a non-thermal distribution of the 
decay products. The subsequent non-linear 
evolution may end up in thermalizing the content of the universe.

%through a decay 
%width was improved 
%Energy transfer during 
%coherent oscillations regime 
%can be very efficient if the 
%fields' modes enter Broad Parametric Resonance (BPR) regime. 
%Particle production and reheating temperature may be thus 
%affected by this phenomenon. Inflaton modes themselves and 
%the perturbations generated during inflation can be, in principle, 
%enhanced by BPR giving rise to observable effects in the CMBR.\\

In this section we study the effect of the parametric resonance for the 
various types of fluctuations present and coupled to the inflaton during 
its coherent oscillations. For the analytic treatment we use Eq. (\ref{coherent}) for the evolution of the background, and check our results by numerical analysis.
For the study of fluctuations during the coherent oscillations regime
the cosmic time is useful and one rewrites, through a suitable change of variable, the equations of motion of the diverse fields in a Mathieu-like form:
\be{matform}
\frac{d^{2} y (t)}{d \pa{\Omega \,t}^{2}} + \left[ A(t) - 2 q (t) \sin 2 \Omega t \right] y (t) = 0
\ee
where $\Omega$ is a frequency which may depend on the field considered.
We call the above equation a Mathieu-like equation since in the Mathieu equation $A$ and $q$ do not depend on time. On taking into account the expansion of the universe $A$ and $q$ depend on time and two principal effects arise: $q$ decays in time and fluctuations, characterized by a comoving momentum $k$, move in the $(q,A)$ plane (see for example \cite{Bender}). An examination of the $(q, A)$ plot of the stability/instability regions is useful in order to understand if the (time-)evolution of the modes end in a stage in which inflaton oscillations cannot be neglected.
%\begin{figure}[t!]
%\centering
%\epsfig{file=qAplotFINAL.eps, width=7.5 cm}
%\caption{{\it An example of a (q,A) plot for the Mathieu equation in which the white regions are associated with a stable solution.} 
%\label{qAplot}}
%\end{figure}

The broad resonance regime ($q\gg1$) has been dubbed {\em stochastic resonance} in an expanding universe \cite{Kofman:1997yn}. The intermediate ($q \sim 1$) and narrow ($q\ll1$) resonance regimes are difficult to treat
and the exponential growth which one obtains with $A$ and $q$ constant in time can be completely washed out by the expansion of the universe. Let us remember that the small $q$
region of the first instability band is:
\be{firstband}
1 - q - \frac{q^2}{8} - {\cal O} (q^3) < A < 1 + q + \frac{q^2}{8} + {\cal O} (q^3) \,.
\ee
which is situated around $A \simeq 1$. When $q$ is not large
the effect of oscillations for $y$ arises mainly if the fluctuation spends time in the first resonant band.

%The general approach consists in rewriting, through a suitable change of variable, the equations of motion of the diverse fields in a Mathieu-like form:
%\be{mathieueq}
%\ddot y(t)+\paq{1-2 q\sin 2t}y(t)=0
%\ee
%where the dot in the above denotes the derivative with respect to $t$, and $q$ is constant. For $2\, q < 1$ this equation leads to a narrow parametric resonance and for $2\, q> 1$ it leads to a broad parametric resonance. On considering the first instability band we find that it is the former case that occurs.

\subsection{Gravitational Waves}
We begin by considering gravitational waves.
The equation governing the dynamics of tensor perturbations (\ref{tnperteqN}), on setting $\tilde h_{k}\equiv a^{3/2}\sigma\,h_{k}$,
can be cast in the form of the Mathieu like equation
\be{eqtenPRE}
\frac{\rd^{2} \tilde h_{k}}{\rd t^{2}}+m_{\rm eff,t}\pa{t}^{2} \tilde h_{k}=0
\ee
and the effective mass is
\be{mefften}
m_{\rm eff,t}\pa{t}^{2}=\omega_{0}^{2}\pa{\frac{k^{2}}{a^{2}\omega_{0}^{2}}
+\frac{f\,r}{1+r\,\omega_{0}\,t}\sin\omega_{0}t} +\mathcal{O}\pa{\frac{1}{t^{2}}}\,.
\ee
For $r \omega_0 t \gg 1$ this equation can be rewritten as 
\be{hPRE}
\frac{d^2 \tilde{h_k}}{d (\omega_{0} t/2)^2} + \left[ A_h + 
2 q_{h} \sin \left( \omega_0 t \right) \right] \tilde{h_k} = 0
\ee
where $A_h$ and $q_h$ are time dependent functions
\be{hPREpar}
A_h (t) =\frac{4 k^2}{a^2 \omega_{0}^2}\,,\quad q_{h} = \frac{f}{2 \omega_{0} t}\,
\ee
and we omit terms of order $1/t^{2}$.
For $r \omega_0 t \gg 1$ the trajectory in the $(q,A)$ plane is therefore $A_h (t) \propto 
k^2 q_h^{4/3} (t)$.
The wavelengths pass through the first resonance band but end in the stability region.

%and no BPR regime occur in the gravitational waves sector.\\

\subsection{Inflaton Fluctuations}

%If one considers inflaton modes dynamics for a generic, potential 
%(with a non trivial minimum) then t

On considering inflaton mode dynamics for a generic potential (with a non trivial minimum) Eq.\ (\ref{scperteq}) can be recast in the form of a Mathieu like equation by rescaling  $\tilde{\delta \sigma_k} = \sqrt{a^3 Z} \delta \sigma_k$ and keeping contributions of the homogeneous mode fluctuations $\ds$ to first order. The final equation for $\tilde{\ds_{k}}$ turns out to be
\be{eqinfPRE}
\frac{\rd^{2}\tilde{\ds_{k}}}{\rd  t^{2}}
+m_{\rm eff \,, \sigma}\pa{t}^{2}\,\tilde{\ds_{k}}=0
\ee
where
\begin{eqnarray}\label{meffinf}
m_{\rm eff,s}\pa{t}^{2}\equiv&&\omega_{0}^{2}
\paq{1-\frac{9\,f\,r}{1+r \,\omega_{0}\,t}\cos\omega_{0}t\right.\nonumber\\
&&\left.\times\pa{1+\frac{4\sqrt{6\gamma\pa{1+6\gamma}}}{9\gamma}\sin\omega_{0}t}}\nonumber\\
&&+\mathcal{O}\pa{\frac{1}{t^{2}}}
\end{eqnarray}
and $f$ and $r$ are defined in (\ref{defMSA}). Thus, in the $r\,\omega_{0}\,t\gg1$ regime, 
Eq.\ (\ref{eqinfPRE}) reduces to
\be{eqinfPRE2}
\frac{d^2 \tilde{\delta \sigma_k}}{d (\omega_{0} t)^2} + \left[ A_\sigma + 
2 q_{\sigma \, 1} \sin \left( 2 \omega_{0} t \right)
+ 2 q_{\sigma \, 2} \sin \left( \omega_{0} t \right) \right] \tilde{\delta \sigma_k} = 0
\ee
where $A_{\sigma}$, $q_{\sigma\,1}$ and $q_{\sigma\,2}$ are time dependent functions of the form 
\be{infPREdef}
A_\sigma =\frac{ k^2}{a^2 \omega_{0}^2} + 1\,,\quad q_{\sigma \, 1} = \frac{2}{\omega_{0} t}\,,
\quad q_{\sigma \, 2} = \frac{\sqrt{\frac{27 \gamma}{2 (1+6\gamma)}}}{\omega_{0} t}
\ee
and we omit terms of order $1/t^{2}$.
The equation for the rescaled IG inflaton fluctuations therefore exhibits two oscillating terms: whereas the one multiplied by $q_{\sigma\,2}$ is similar to the term appearing for gravitational waves, $q_{\sigma\,1}$ does not depend on $\gamma$. On neglecting $q_{\sigma\,2}$ one has the following trajectory in the $(q,A)$ plane:
\be{mathieuscalar}
A_\sigma (t) = 1 + \frac{k^2}{a^2 \omega_{0}^2} = 1 + \frac{k^2}{a_{0}^{2}\omega_{0}^2}
\left( \frac{\omega_{0} t_0}{2} \right)^{4/3} q_{\sigma \, 1}^{4/3} (t) \,,
\ee
which shows how fluctuations end in the first resonance band asymptotically. The fluctuation $\delta\tilde {\sigma}_k$ grows linearly in time as shown carefully by the MSA method in the Appendix. This means that the inflaton fluctuations oscillate with constant amplitude [9],
rather than decay if oscillations in the homogeneous inflaton are averaged in time.
Since $q_{\sigma\,1}$ does not depend on $\gamma$ our analysis applies to the coherent oscillation regime of chaotic quadratic inflation in Einstein gravity as well, in agreement with Ref. \cite{Nambu:1996gf}. Therefore during the coherent oscillations of a massive inflaton, gauge-invariant inflaton fluctuations oscillate with constant amplitude on large scales \cite{Finelli:1998bu} and also on small scales.
We have numerically checked that the inclusion of the oscillation term multiplied by $q_{\sigma\,2}$ does not modify the qualitative behaviour just described. Indeed the amplitude of this term is never important, $q_{\sigma\,2} \sim {\cal O} (\gamma/(\omega_0 t))$ for $\gamma \ll1$ and $q_{\sigma\,2} \sim 3/(2 \omega_0 t)$ for $\gamma \gg 1$, and the frequency of oscillations is half of the dominant one.

%The equation for the rescaled IG inflaton fluctuations therefore exhibits two oscillating terms: whereas the term multiplied by 
%$q_{\sigma\,2}$ is similar to the one appearing for gravitational waves, $q_{\sigma\,1}$ does not depend on $\gamma$. 
%For $a$ sufficiently large, on neglecting $q_{\sigma\,2}$, $A_{\sigma}$ becomes of order $1$, one winds up in the first instability region and has narrow parametric resonance.

%BPR would occurs for $A$ close enough to half the frequencies of oscillation 
%in (\ref{eqinfPRE2}), namely for $A\sim 1/2$ and $A \sim 1$, thus only the 
%oscillating term with amplitude $q_{1}$ contributes to a BPR behavior. 
%When the wavelength of the inflaton modes is long compared with $\omega_{0}^{-1}$ 
%one observes a linear increase of the amplitude $\tilde{\delta \sigma_k}\sim t$ 
%(see Appendix I for details).\\

\subsection{Scalar Fields non-minimally coupled to gravity}

We finally take into consideration the evolution of a scalar test field $\chi$ non-minimally 
coupled to gravity and interacting with the inflaton. Let
\be{actionSF}
S_{\chi}=\int \rd x^{4}\sqrt{-g}\paq{-\frac{g^{\mu\nu}}{2}\partial_{\mu}\chi\partial_{\nu}\chi-\frac{m_{\chi}^{2}}{2}\chi^{2}+\mathcal{L}_{\rm int}}
\ee
be the action for such a field and
\be{chiinter}
\mathcal{L}_{\rm int}=-\frac{\xi}{2}R\chi^{2}+\frac{g^{2}}{2}\sigma^{2}\chi^{2}
\ee
the interaction with $\xi>0$. On considering only the interaction of the modes of the field $\chi$ with the homogeneous part of the inflaton one finally obtains
\be{eqchiPRE}
\frac{\rd^{2} \tilde \chi_{k}}{\rd t^{2}}+m_{\rm eff,\chi}\pa{t}^{2} \tilde \chi_{k}=0
\ee
where
\begin{eqnarray}
m_{\rm eff,\chi}\pa{t}^{2}=\!\!\!&&\omega_{0}^{2}\left[\frac{k^{2}}{a^{2}\omega_{0}^{2}}+\frac{3\pa{4\xi-1}\,f\,r}{1+r\,\omega_{0}\,t}\sin\omega_{0}t\right.\nonumber\\
&&+2g^{2}\frac{1+6\gamma}{\mu}\frac{f\,r}{1+r\,\omega_{0}\,t}\sin\omega_{0}t\nonumber\\
&&\left.+\frac{m_{\chi}^{2}}{\omega_{0}^{2}}+g^{2}\frac{1+6\gamma}{2\mu}\right]+\mathcal{O}\pa{\frac{1}{t^{2}}}\label{meffchi}
\end{eqnarray}
and $\tilde\chi_{k}=a^{3/2}\chi_{k}$. From (\ref{meffchi}) we see that the scalar field does not end up in the instability region unless 
\be{BRchi}
\frac{m_{\chi}^{2}}{\omega_{0}^{2}}+g^{2}\frac{1+6\gamma}{2\mu}\simeq \frac{1}{2},
\ee
which shows how a mass $m^{2}_\chi$ generically suppresses the resonance.
Note also that a broad resonance regime is possible for $m_\chi=g=0$ for large $\xi$.
%Let us end this section by observing that some concern has been raised about the evolution of inflation and preheating because of a possible
%tachyonic instability due to the potential for small field inflation \cite{Felder,LindeNew}.
%In our previous manuscript we observed that, in the SF regime, for sufficient inflation a small shift of the inflaton from the minimum appears to be sufficient. In particular for small enough $\gamma$ one can actually reduce even further the shift necessary and thus avoid the problem. We hope to examine this point in the future.

\section{Conclusions}
We have investigated in detail inflation in the IG framework where inflation is driven by the same scalar field as that associated with the observed value of Newton's constant. Inflation in IG can be successful in the SR regime, and leads to a nearly flat spectrum of scalar perturbations and a small tensor to scalar ratio compatibly with observations. The SR conditions are not simply associated with the shape of the inflaton potential since the dynamics of the scalar field strongly depends on its coupling $\gamma$ to gravity. We have made accurate comparisons with observations for different ``symmetry breaking'' potentials and showed how their shape is relevant for successful predictions of the model. In particular we have seen that the LG and CW type potentials are compatible with the data for large field inflation and for suitable values of the parameters for small field inflation. For potentials involving higher powers of the scalar field (with respect to the previously mentioned cases) agreement with the data is more problematic and if it occurs one obtains constraints on $\gamma$ for both the large and small field cases.

We have also studied, always in IG, the coherent oscillatory regime and obtained an accurate analytical solution for the dynamics of inflaton and of the Hubble parameter by using a MSA. We employed the solution to first investigate the behavior of the average evolution of the Universe. We found that, independently of the average equation of state of the scalar field, the Universe expands as $\av{H(t)}\propto2/(3t)$.\\ 
Further perturbative and resonant reheating have also been studied. We found, in the former case, that the decay of the inflaton into ordinary matter is less efficient for $\gamma$ large and that the reheating temperature is comparable with the EG value for $\gamma$ small. We also examined the parametric amplification of the scalar and tensor perturbations and of a generic scalar field non-minimally coupled to gravity and coupled to the inflaton. We note that parametric resonance occurs for scalar perturbations with $H\ll k/a\ll\omega_{0}$ and such perturbations maintain a constant amplitude instead of decaying as the Universe expands. Finally parametric resonance does not occur for gravitational waves whereas amplification may occur for a generic scalar field depending on the parameters of its Lagrangian. 

%%%%%%%%%%%%%%%%%%%%
\appendix
\section{Preheating}
In an expanding spacetime, parametric resonance in the first instability band occurs when the Mathieu-like equation governing modes dynamics can be cast in the form 
\be{BPRdyn}
\ddot y+\paq{1+\epsilon\, q\pa{t}\sin2\,t}y=0
\ee
where $\epsilon$ is a small dimensionless parameter. The theory of parametric resonance is well known when $q\pa{t}={\rm const}$ but in an expanding spacetime, where a time dependence generally appears, the dynamics leads to quite different results. MSA gives the correct approximate behavior for (\ref{BPRdyn}).\\
Since $\epsilon\,q\pa{t}\ll1$ one can look for an approximate solution in the form
\be{appsolAPP}
y=y_{0}\pa{t,\tau}+\epsilon\,y_{1}\pa{t,\tau}
\ee
where $\tau\equiv \epsilon\,t$. On comparing orders one has
\begin{eqnarray}
\!\!\!\!\!\!\!\!&&\frac{\partial^{2}y_{0}}{\partial t^{2}}+y_{0}=0\Longrightarrow y_{0}=A\pa{\tau}\,\re^{i\,t}+{\rm c.c.} \\
\!\!\!\!\!\!\!\!&&\frac{\partial^{2}y_{1}}{\partial t^{2}}+y_{1}+2\frac{\partial^{2}y_{0}}{\partial t\partial \tau}+q\pa{t}\frac{\re^{2\,i\,t}-\re^{-2\,i\,t}}{2\,i}y_{0}=0\label{order1APP}
\end{eqnarray}
and the requirement that secular contributions cancel in (\ref{order1APP}) leads to the differential equation 
\be{eqAAPP}
4\frac{\rd A}{\rd t}-q\pa{t}A^{*}=0.
\ee
On separating $A$ into real and imaginary parts ($A=B+i\,C$) growing and decaying modes also separate
\begin{eqnarray}
4\frac{\rd B}{\rd t}-q\pa{t}B&=&0\\
4\frac{\rd C}{\rd t}+q\pa{t}C&=&0
\end{eqnarray}
and, if $q\pa{t}>0$, the imaginary part decays.\\
In general, the expansion of the universe during reheating leads to $q\pa{t}\sim p/t^{n}$ and in this case the growing dynamics becomes
\be{solAPP}
B\pa{t}=B_{0}\exp\int_{t_{0}}^{t}\frac{p}{4\,\tilde t^{n}}\rd\tilde t.
\ee
When $n>1$ the amplitude of oscillations increases but still exhibits a constant asymptotic behaviour $B\sim B_{0}\exp\paq{{\frac{p}{4\pa{n-1}}t_{0}^{1-n}}}$, for $n<1$ the amplitude increases exponentially as $B\sim B_{0}\exp\paq{\frac{p}{4\pa{1-n}}t^{1-n}}$. In contrast, when $n=1$ (which is the value we find for scalar perturbations in the instability region), one obtains the following power-law dependence
\be{PLsolAPP}
B=B_{0}\pa{\frac{t}{t_{0}}}^{p/4}
\ee
which, for $p=4$ corresponds to $\delta\tilde\sigma\sim t$, as mentioned in Section VIIB.


\begin{thebibliography}{99}

\bibitem{brans}
  C.~Brans and R.~H.~Dicke,
  %``Mach's principle and a relativistic theory of gravitation,''
  Phys.\ Rev.\  {\bf 124} (1961) 925.
  %%CITATION = PHRVA,124,925;%%
  
 \bibitem{sakharov}
A. D. Sakharov, Dokl. Akad. Nauk. SSSR 117, 70 (1967); 
[Sov. Phys. Dokl. 12, 1040 (1967)].



%\cite{Zee:1980sj}
\bibitem{Zee:1980sj}
  A.~Zee,
  %``Spontaneously Generated Gravity,''
  Phys.\ Rev.\  D {\bf 23}, 858 (1981).
  %%CITATION = PHRVA,D23,858;%%

\bibitem{adler}
S. Adler, Rev. Mod. Phys. 54, 729 (1982)

\bibitem{CV}
F.~Cooper and G.~Venturi,
Phys.\ Rev.\  D {\bf 24} (1981) 3338.
%%CITATION = PHRVA,D24,3338;%%

\bibitem{zee}
A. Zee, Phys. Rev. Lett. 42, 417 (1979)

\bibitem{Accetta}
  F.~S.~Accetta, D.~J.~Zoller and M.~S.~Turner,
  %``Induced Gravity Inflation,''
  Phys.\ Rev.\  D {\bf 31} (1985) 3046.
  %%CITATION = PHRVA,D31,3046;%%
  
\bibitem{Sorbo}
  N.~Kaloper, L.~Sorbo and J.~Yokoyama,
  %``Higgsflation at the GUT scale in a Higgsless Universe,''
  Phys.\ Rev.\  D {\bf 78} (2008) 043527
  [arXiv:0803.3809 [hep-ph]].
  %%CITATION = PHRVA,D78,043527;%%

\bibitem{CW}
S.~R.~Coleman and E.~J.~Weinberg, Phys.\ Rev.\  D {\bf 7} (1973) 1888.
%%CITATION = PHRVA,D7,1888;%%

\bibitem{spokoiny}
B. L. Spokoiny, Phys. Lett. 147B, 39 (1984).

\bibitem{FTV}
  F.~Finelli, A.~Tronconi and G.~Venturi,
  %``Dark Energy, Induced Gravity and Broken Scale Invariance,''
  Phys.\ Lett.\  B {\bf 659} (2008) 466
  %%CITATION = PHLTA,B659,466;%%

\bibitem{CFTV}
A. Cerioni, F.~Finelli, A.~Tronconi and G.~Venturi, 
Phys. Lett. B 681 (2009) 383.
%%CITATION = PHLTA,B681,383;%%

\bibitem{Maeda}
  K.~i.~Maeda,
  %``Towards the Einstein-Hilbert Action via Conformal Transformation,''
  Phys.\ Rev.\  D {\bf 39} (1989) 3159.
  %%CITATION = PHRVA,D39,3159;%%

\bibitem{Flanagan}
  E.~E.~Flanagan,
  %``The conformal frame freedom in theories of gravitation,''
  Class.\ Quant.\ Grav.\  {\bf 21} (2004) 3817
  [arXiv:gr-qc/0403063].
  %%CITATION = CQGRD,21,3817;%%
  
\bibitem{Chiba}
  T.~Chiba and M.~Yamaguchi,
  %``Extended Slow-Roll Conditions and Rapid-Roll Conditions,''
  JCAP {\bf 0810} (2008) 021
  [arXiv:0807.4965 [astro-ph]].
  %%CITATION = JCAPA,0810,021;%%

\bibitem{hwang}
J.~c.~Hwang, Class.\ Quant.\ Grav.\  {\bf 14} (1997) 3327
%%CITATION = CQGRD,14,3327;%%

\bibitem{fhll}
  F.~Finelli, J.~Hamann, S.~M.~Leach and J.~Lesgourgues,
  %``Single-field inflation constraints from CMB and SDSS data,''
  JCAP {\bf 1004} (2010) 011
  [arXiv:0912.0522 [Unknown]].
  %%CITATION = JCAPA,1004,011;%%
  
\bibitem{turner}
  M.~S.~Turner,
  %``Coherent Scalar Field Oscillations In An Expanding Universe,''
  Phys.\ Rev.\  D {\bf 28}, 1243 (1983).
  %%CITATION = PHRVA,D28,1243;%%

\bibitem{TB}
  J.~H.~Traschen and R.~H.~Brandenberger,
  %``PARTICLE PRODUCTION DURING OUT-OF-EQUILIBRIUM PHASE TRANSITIONS,''
  Phys.\ Rev.\  D {\bf 42} (1990) 2491.
  %%CITATION = PHRVA,D42,2491;%%

\bibitem{Kofman:1997yn}
 L.~Kofman, A.~D.~Linde and A.~A.~Starobinsky,
 %``Towards the theory of reheating after inflation,''
 Phys.\ Rev.\  D {\bf 56} (1997) 3258
 [arXiv:hep-ph/9704452].
 %%CITATION = PHRVA,D56,3258;%%

\bibitem{KLS}
  L.~Kofman, A.~D.~Linde and A.~A.~Starobinsky,
  %``Reheating after inflation,''
  Phys.\ Rev.\ Lett.\  {\bf 73} (1994) 3195
  [arXiv:hep-th/9405187].
  %%CITATION = PRLTA,73,3195;%%

\bibitem{Nambu:1996gf}
 Y.~Nambu and A.~Taruya,
 %``Evolution of cosmological perturbation in reheating phase of the
 %universe,''
 Prog.\ Theor.\ Phys.\  {\bf 97} (1997) 83
 [arXiv:gr-qc/9609029].
 %%CITATION = PTPKA,97,83;%%

\bibitem{Finelli:1998bu}
 F.~Finelli and R.~H.~Brandenberger,
 %``Parametric amplification of gravitational fluctuations during  reheating,''
 Phys.\ Rev.\ Lett.\  {\bf 82} (1999) 1362
 [arXiv:hep-ph/9809490].
 %%CITATION = PRLTA,82,1362;%%

%\bibitem{nambutaruya}
  %Y.~Nambu and A.~Taruya,
  %``Evolution of cosmological perturbation in reheating phase of the
  %universe,''
  %Prog.\ Theor.\ Phys.\  {\bf 97} (1997) 83
  %%CITATION = PTPKA,97,83;%%


%\bibitem{Felder}
  %G.~N.~Felder, J.~Garcia-Bellido, P.~B.~Greene, L.~Kofman, A.~D.~Linde and I.~Tkachev,
  %``Dynamics of symmetry breaking and tachyonic preheating,''
  %Phys.\ Rev.\ Lett.\  {\bf 87} (2001) 011601
  %%CITATION = PRLTA,87,011601;%%


%\bibitem{LindeNew}
  %M.~Desroche, G.~N.~Felder, J.~M.~Kratochvil and A.~D.~Linde,
  %``Preheating in new inflation,''
  %Phys.\ Rev.\  D {\bf 71} (2005) 103516
  %%CITATION = PHRVA,D71,103516;%%

%\cite{Komatsu:2008hk}
\bibitem{Komatsu:2008hk}
  E.~Komatsu {\it et al.}  [WMAP Collaboration],
  %``Five-Year Wilkinson Microwave Anisotropy Probe (WMAP\altaffilmark 1 )
  %Observations:Cosmological Interpretation,''
  Astrophys.\ J.\ Suppl.\  {\bf 180}, 330 (2009)
  [arXiv:0803.0547 [astro-ph]].
  %%CITATION = APJSA,180,330;%%
  
 \bibitem{Bender}
  C.~M.~Bender and S.~A.~Orszag,
  \emph{Advanced mathematical methods for scientists and engineers: Asymptotic methods and perturbation theory}
  (Springer Verlag, 1999)


\end{thebibliography}
\end{document}